\documentclass[journal]{IEEEtran}

\usepackage{float}
\usepackage{amsfonts,amsopn,dsfont} 
\usepackage{color}
\usepackage{amsmath}
\usepackage[table]{xcolor}
\usepackage{graphicx,epstopdf}
\usepackage{algorithm}
\usepackage{algorithmic}
\usepackage{cite}
\usepackage{caption}
\usepackage{subcaption}
\usepackage[normalem]{ulem}
\usepackage[colorlinks = true,
            linkcolor = blue]{hyperref}

\def\bfy{\boldsymbol{y}}
\def\bfs{\boldsymbol{s}}
\def\bfz{\boldsymbol{z}}

\def\btheta{\boldsymbol{\theta}}

\def\mV{\mathcal{V}}

\newenvironment{algogo}[1]{
\smallskip
\noindent \hrule\vspace{0.2\baselineskip} \hrule
\begin{small}
\refstepcounter{algo} \center{\bf \textsc{Algorithm \thealgo}}
\\{\center{\bf #1}}
\smallskip
\flushleft
 } {
\end{small}
\smallskip
\hrule\vspace{0.2\baselineskip} \hrule
\smallskip
}
\newcounter{algo}
\renewcommand{\thealgo}{\arabic{algo}}

\begin{document}

\title{Robust 3D reconstruction of dynamic scenes from single-photon lidar using Beta-divergences}

\author{
	Quentin Legros, Julian Tachella, Rachael Tobin, Aongus McCarthy,~\IEEEmembership{Member,~IEEE,} Sylvain Meignen, Gerald S. Buller, Yoann~Altmann,~\IEEEmembership{Member,~IEEE,}
	Stephen~McLaughlin,~\IEEEmembership{Fellow,~IEEE}
	and~Michael~E.~Davies,~\IEEEmembership{Fellow,~IEEE}
	\thanks{}}

\maketitle

\begin{abstract}
In this paper, we present a new algorithm for fast, online 3D reconstruction of dynamic scenes using times of arrival of photons recorded by single-photon detector arrays. 
One of the main challenges in 3D imaging using single-photon lidar in practical applications is the presence of strong ambient illumination which corrupts the data and can jeopardize the detection of peaks/surface in the signals. This background noise not only complicates the observation model classically used for 3D reconstruction but also the estimation procedure which requires iterative methods. In this work, we consider a new similarity measure for robust depth estimation, which allows us to use a simple observation model and a non-iterative estimation procedure while being robust to mis-specification of the background illumination model.
This choice leads to a computationally attractive depth estimation procedure without significant degradation of the reconstruction performance. This new depth estimation procedure is coupled with a spatio-temporal model to capture the natural correlation between neighboring pixels and successive frames for dynamic scene analysis. The resulting online inference process is scalable and well suited for parallel implementation. The benefits of the proposed method are demonstrated through a series of experiments conducted with simulated and real single-photon lidar videos, allowing the analysis of dynamic scenes at 325 m observed under extreme ambient illumination conditions. 
\end{abstract}


%
\IEEEpeerreviewmaketitle
\section{Introduction}
\label{sec:intro}
Fast and reliable reconstruction of 3D scenes using single-photon light detection and ranging (lidar) is extremely important for a variety of applications, including environmental monitoring \cite{mallet2009full,Canutoeaau0137}, autonomous driving \cite{hecht2018lidar} and defence \cite{gao2011research,Tobin:19}. While 3D profiles can be obtained from a range of modalities, single-photon lidar (SPL) offers appealing advantages, including low-power imaging, a capability for long-range imaging \cite{Pawlikowska:17,Li2019_arxiv,Li2020} or imaging in complex media such as fog/smoke \cite{Tobin:19} and turbid underwater environments\cite{maccarone2015underwater} with excellent range resolution (of the order of millimetres \cite{McCarthy:09}). 

Over the last few years, a wide range of reconstruction algorithms has been proposed to reconstruct individual depth images from SPL data, e.g., \cite{shin2015photon,shin2016photon,altmann2018eusipco,Lindell:2018:3D,rappfew,Tachella2019_manipop,Rapp2019,Chan2019}. Several algorithms have also been proposed to analyze distributed objects \cite{shin2016computational,tobin2017long,halimi2017,Tachella2019_manipop,Tachella2019_ICASSP,Tachella2019_RT3D}, i.e., when multiple surfaces are visible within each pixel. Irrespective of the number of surfaces visible in each pixel, one of the main goals of these algorithms is to reconstruct high quality depth profiles from as small a photon budget as possible (see also \cite{McCarthy:09,altmann2016lidar,altmann2016target,halimi2016restoration}) and it was shown that reconstruction from as few as one photon per pixel is possible under favorable observation conditions. Since single-photon lidar technology consists of illuminating the scene with a pulsed laser and analyzing the time of arrival (ToA) of reflected photons, successful reconstruction from a few return photons enables the consideration of shorter integration/acquisition times and thus the analysis of highly dynamic scenes. Note that most existing methods are to be used offline since the computational time required to reconstruct a point cloud is usually longer than the acquisition time allocated for a single frame. However, a recent study has presented results of the reconstruction of complex scenes at video frame rates \cite{Tachella2019_RT3D}.

A shared property of all the algorithms mentioned above is that they concentrate on the reconstruction of one point cloud per time frame, processing a video as a sequence of independent frames. While a method was recently proposed in \cite{Halimi2020} to jointly process batches of SPL frames, it remains computationally intractable for long video sequences due to memory requirements. Thus, there is a clear need for scalable and reliable methods able to adaptively process the increasing amount of single-photon data recorded by new single-photon avalanche diode (SPAD) detector arrays \cite{Ren:18,Henderson2019}, offering a growing number of pixels. In \cite{Altmann2019}, we proposed an online reconstruction method, relying on individual photon-detection events, e.g., binary frames, which was used to reconstruct sequentially a series of depth images using at most one photon per pixel and frame.
Although this sequential approach leverages correlations between successive time frames, the method does not allow the analysis of histograms (i.e., reconstruction after several illumination periods), as the estimation is performed after each illumination period.
Moreover, this method is not adapted to situations where the ambient illumination levels lead to low signal-to-background ratio (in particular, smaller than one). 

In this paper, we consider lidar data acquired using SPAD arrays and investigate a new 3D reconstruction algorithm that accounts for the temporal correlation between successive point clouds to be reconstructed. 
More precisely, we address the problem of the reconstruction of a temporal series of point clouds, where each point cloud is associated with a different integration period, over which the pulsed laser emits an arbitrary number of pulses. This integration period is user-defined and we assume that the scene is quasi-static during that period. 
In contrast to \cite{Altmann2019}, the sequential reconstruction by the proposed algorithm is performed after an arbitrary number of periods and thus does not make any assumptions regarding the number of detection events to be used in the reconstruction of each point cloud.
The most basic and fastest method to estimate the distance of an object from a SPL histogram is via log-matched filtering but this method fails if the background level due to ambient illumination and light scattering is too high. In such cases where the background cannot be neglected, it is traditionally included in the observation model and is estimated to improve the depth estimation. However, the resulting model makes the estimation process more complicated and slower iterative schemes (optimization-based \cite{shin2016computational} or simulation-based \cite{altmann2016lidar}) are classically used. Strong ambient levels are encountered in many practical applications, for instance in long-range imaging applications in free-space \cite{Pawlikowska:17,Li2019_arxiv,Li2020} and challenging imaging applications through scattering media such as turbid underwater environments \cite{maccarone2015underwater,halimiwater,Maccarone:19} and through fog/smoke \cite{Tobin:19}. It is thus extremely important to develop methods adapted to such challenging observation conditions.

In contrast to existing reconstruction methods, we propose a depth estimation method that does not require the background level to be modeled while allowing robust depth estimation when the background cannot be neglected. 
Instead of defining a likelihood function based on an observation model assumed fully specified and accurate, we define a pseudo-likelihood which only depends on the target depth. A robust estimation procedure is then developed to account for the mismatch between the simplified observation model and the actual data distribution. More precisely, the proposed data fidelity term is based on a $\beta$-divergence instead of the classical Kullback-Leibler divergence, allowing efficient reconstruction performance in the presence of an unknown (and high) ambient illumination level. Adopting a Bayesian approach, this pseudo-likelihood is coupled, for each pixel and each frame, with a depth prior model to derive the so-called pseudo-posterior distribution of the position of each surface. This distribution is then used to perform surface detection (i.e., deciding if a surface is actually visible or not) and to define the prior distribution of the point cloud in the next frame, in an adaptive fashion.

The main contributions of this work are:
\begin{itemize}
\item A new pseudo-Bayesian model for robust 3D reconstruction using streams of photon detection events in the presence of high ambient illumination levels
\item An efficient depth estimation strategy which presents a fixed and predictable computational cost (in contrast to most existing methods which are iterative and which require a convergence criterion to be reached). 
\item A new online/sequential estimation strategy, proposed to the best of our knowledge for the first time, for reconstruction of dynamic 3D scenes from streams of photon detection events. This method based on assumed density filtering is highly scalable and computationally attractive. It also includes an automatic and principled surface detection method originally proposed in \cite{TachellaEUSIPCO2019}, included for the first time in a sequential reconstruction process.
\item{A re-interpretation of the classical matched-filtering approach adopted for depth estimation using SPL as a robust estimator.}
\end{itemize}

The remainder of the paper is organized as follows. Section \ref{sec:robust} recalls the classical observation models for 3D reconstruction using SPL measurements in the photon-starved regime, introduces the similarity measure based on $\beta$-divergences for robust estimation, and demonstrates its benefits for the analysis of a single frame. The spatio-temporal model and new online reconstruction method are detailed in Section \ref{sec:online}. Results of simulations conducted with real and synthetic sequences of frames/histograms are presented and discussed in Section \ref{sec:results}. 
Conclusions are finally reported in Section \ref{sec:conclusion}.

\section{Robust depth estimation}
\label{sec:robust}
\subsection{Observation models}
\label{subsec:likelihood}
In this work, we consider a sequence of $N$ temporal frames which consist of $P$ pixels. More precisely, for each frame, the data associated with each pixel consists of a set of photon ToAs. This paper addresses the reconstruction of dynamic 3D scenes where the $N$ frames are processed sequentially to reduce data storage requirements and account for the temporal correlation between successive frames. 
In this section, we discuss observation models and estimation strategies for the analysis of a pixel of a single frame. For now, we assume that one surface is visible in each pixel and we do not introduce pixel/frame indices in this section to keep notation clear.

\subsubsection{Ideal model}
Assuming that the ambient illumination and detector dark counts can be neglected, the recorded ToAs are only associated with photons originally emitted by the laser source. For a given pixel, the probability density function of a photon ToA $y \in (0,T)$, where $T$ is the repetition period of the laser source, is given by 
\begin{eqnarray}
\label{eq:lik_ideal}
f_0(y| d ) = s_0\left(y - \frac{2 d}{c}\right),
\end{eqnarray} 
where $d$ is the distance between the imaging system and the surface of interest and $c$ is the light speed in the medium.
In \eqref{eq:lik_ideal}, we assume that the scene is approximately static (within each frame) and $s_0\left(\cdot\right)$ is the normalized instrumental response function (IRF) of the lidar system which can be measured during calibration of the imaging system. To simplify notation, we assume that $s_0\left(\cdot\right)$ is the same for all the pixels but the method proposed also applies if the shape of the IRF is pixel-dependent. Note that we also assume that the shape of $s_0\left(\cdot\right)$ remains the same for all the admissible values of $d$.

When $K$ photons are detected, if the dead-time of the detector can be neglected, the photon ToAs are mutually independent (given $d$) and the joint likelihood is given by $f(\bfy | d)=\prod_{k} f_0(y_{k}|d)$ with $\bfy =\{y_{k} \}_k$. 
Due to the finite timing resolution of SPAD detectors, the recorded ToAs are not continuous variables but instead live on a grid, whose resolution depends on the system used. Thus, the photon-starved regime it is possible to consider a model equivalent to \eqref{eq:lik_ideal}, based on Poisson noise and which can be expressed as
\begin{eqnarray}
\label{eq:lik_ideal_poisson}
z_t |r,d \sim \mathcal{P} \left(r \tilde{s}_0\left(t - \frac{2 d}{c}\right)\right), \quad t=1,\ldots, N_T
\end{eqnarray}
where $N_T$ is the number of non-overlapping time bins spanning $(0,T)$, $\tilde{s}_0\left(t - \frac{2 d}{c}\right)=\int_{y=(t-1)T/N_T}^{tT/N_T} s_0\left(y - \frac{2 d}{c}\right) \textrm{d}y$, and $z_t$ is the number of photons detected in the $t$th time bin. In \eqref{eq:lik_ideal_poisson}, $\mathcal{P}(\lambda)$ denotes the Poisson distribution with mean $\lambda$.
Moreover, $r \geq 0$ is an amplitude parameter which mainly depends on the number of laser pulses sent during the frame, the efficiency of the detector and the reflectivity of the object. In this case, we obtain the joint likelihood $f(\bfz |r,d)=\prod_{t} f(z_t|r,d)$, where $\bfz=[z_1,\ldots,z_{N_T}]^T$ is the ToA histogram constructed from $\bfy$ and the maximum likelihood estimator of $d$ can be computed by maximizing the cross-correlation between the logarithm of  $\tilde{\bfs}_{0}=[\tilde{s}_0\left(1 - \frac{2 d}{c}\right),\ldots,\tilde{s}_0\left(T - \frac{2 d}{c}\right)]^T$ and $\bfz$ \cite{altmanndetect2016}.

\subsubsection{Accounting for background detections}
In many practical applications however, the models in \eqref{eq:lik_ideal}-\eqref{eq:lik_ideal_poisson} are not well adapted as background illumination cannot be neglected. Detection events arising from dark counts and additional sources (e.g., solar background) often present a uniform distribution and a more accurate observation model is the following mixture of distributions \cite{altmann2018eusipco}.
\begin{eqnarray}
\label{eq:lik_bg}
f(y | d,w)= w s_0\left(y - \frac{2 d}{c}\right) + (1-w)  \mathcal{U}_{\left(0;T\right)} (y),
\end{eqnarray} 
where $w$ is the pixel and frame dependent probability of a detected photon to be a ``signal'' photon originally emitted by the laser source. Note however that other distributions could be used to account for the nature of the background photons. This probability relates to the signal-to-background ratio (SBR) defined by $\text{SBR}=w/(1-w)$. In that case, estimating $d$ from the joint likelihood $f(\bfy | d,w)=\prod_{k} f(y_{k}  | d,w)$
becomes more challenging since $w$ is usually unknown and needs to be estimated jointly with $d$. Even if $w$ is known, the estimation of $d$ is challenging as the evaluation of $f(\bfy | d,w)$ consists of a product of $K$ mixtures. Similarly, the model in \eqref{eq:lik_ideal_poisson} becomes 
\begin{eqnarray}
\label{eq:lik_bg_poisson}
z_t |r,d \sim \mathcal{P} \left(r \tilde{s}_0\left(t - \frac{2 d}{c}\right) + b \right), \quad t=1,\ldots, N_T,
\end{eqnarray}
where $b \geq 0$ represents the average background level which can be correlated with the target reflectivity $r$. The relationship between $r,b$ and $w$ is obtained through the SBR, i.e., $\text{SBR} =w/(1-w)=r/(b N_T)$.

Although the models described in  \eqref{eq:lik_bg}-\eqref{eq:lik_bg_poisson} are more appropriate than those in \eqref{eq:lik_ideal}-\eqref{eq:lik_ideal_poisson}, they can still fail to describe the data accurately, in particular in scattering media where the distribution of the background photons is not uniform \cite{Maccarone:19}. 
Moreover, the joint estimation of $(w,d)$ or $(r,b,d)$ requires iterative algorithms (e.g., \cite{shin2015photon,altmann2016lidar,Ren:18,altmann2018eusipco}) which can result in a significant computational bottleneck. For these reasons, we investigate robust estimation of $d$ based on \eqref{eq:lik_ideal}, whereby robustness relates to the mismatch between the postulated model and the actual distribution of the ToAs. 

\subsection{Robust estimation using $\beta$-divergences}
Modern methods for depth estimation from single-photon data are statistical methods which use the data likelihood either in a maximum penalized likelihood fashion or within a Bayesian framework\cite{shin2015photon,altmann2016lidar,Ren:18,altmann2018eusipco}. 
Maximum likelihood estimation (MLE) in this context is equivalent to minimizing the Kullback-Leibler (KL) divergence\\
\begin{align}
 D_{\textrm{KL}}(\hat{f}(y)||f(y|\btheta))  = & \int \hat{f}(y) \log \left(\dfrac{\hat{f}(y)}{f(y|\btheta)}\right) \textrm{d} y,\nonumber\\
 = & \textrm{ Const.} -\dfrac{1}{K}\sum_{k}\log\left(f(y_k|\btheta) \right) 
\end{align}
between the empirical distribution of the ToAs, denoted as $\hat{f}(y)=\dfrac{1}{K}\sum_{k}\delta(y-y_k)$, with $\delta(\cdot)$ the Dirac delta function, and the distribution $f(y|\btheta)$, where $\btheta$ is the set of parameters of the postulated model (i.e., $\btheta=d$ if \eqref{eq:lik_ideal} is used, and $\btheta=(d,w)$ if  \eqref{eq:lik_bg} is used). 
Moreover, maximum penalized likelihood approaches can be seen as methods aiming at minimizing a penalized KL divergence (see discussion in Section \ref{subsec:pseudo_post}). 

In this work, we propose to investigate a robust divergence instead of the classical KL divergence when estimating $d$ to account for the mismatch between $\hat{f}(y)$ and the postulated observation model. Moreover, since \eqref{eq:lik_ideal} is simple, i.e., it involves a single parameter, and yields satisfactory results in the low to moderate background regime, it seems reasonable to use this model $f_0(y|d)$ in our 3D imaging strategy instead of \eqref{eq:lik_bg}. 

Among the different families of divergences, we concentrate on $\beta$-divergences defined by\\
$D_{\beta}(g||h)  =  \dfrac{1}{\beta}\int g(y)^{1+\beta}\textrm{d}y$
\begin{eqnarray}
\label{eq:beta_div}
 - \dfrac{\beta+1}{\beta} \int g(y) h(y)^{\beta}\textrm{d}y + \int h(y)^{1+\beta}\textrm{d}y, 
\end{eqnarray}
with $\beta>0$, which generalize the KL divergence \cite{Basu1998}. In addition to being robust to model mismatch (as will be shown in Section \ref{subsec:prelim_results}), this family of divergences allows a computationally attractive estimation of $d$, since the evaluation and minimization of the $\beta$-divergence is simple (e.g., simpler than $\gamma$-divergences \cite{Futami2018}). Note that the KL divergence is asymptotically recovered when $\beta \rightarrow 0$. For our problem, we obtain 
\begin{eqnarray}
\label{eq:beta_div2}
D_{\beta}(\hat{f}(y)||f_0(y|d)) & = & \textrm{Const.} -\dfrac{\beta+1}{\beta K}\sum_{k}f_0(y_k|d)^{\beta}
\end{eqnarray}
where the constant (which depends on $\beta$) does not depend on $d$ since we assume that the shape and the integral of $s_0(\cdot)$ does not depend on $d$, i.e., $\int f_0(y|d)^{1+\beta}\textrm{d}y$ does not depend on $d$, for any $d$ in its admissible set. The performance of any depth estimator based on this $\beta$-divergence depends on the value of the divergence parameter $\beta$. Its impact on the proposed method will be discussed in Section \ref{subsec:prelim_results}.

An interesting result is that using the histogram $\bfz$ (defined in the paragraph below \eqref{eq:lik_ideal_poisson}) instead of $\bfy$ in \eqref{eq:beta_div2}, i.e., using discretized ToAs, minimizing the $\beta$-divergence reduces to maximizing the cross-correlation between $\bfz$ and $\tilde{\bfs}_{0}^{\beta}$, where the exponential function is applied element-wise. Indeed, the sum on the left-hand side of \eqref{eq:beta_div2} becomes $\bfz^T \bfs_{d}^{\beta}$.
In particular, when $\beta=1$, the resulting estimator reduces to the depth estimator obtained via matched filtering. 
Matched filtering is a classical method for peak localization, and is optimal in the presence of white Gaussian noise. Nonetheless, it has been widely used in single-photon lidar analysis \cite{McCarthy2013} and it has been shown empirically to provide similar or even better results than log-matched filtering. 
In this paper, we focus on Bayesian estimation of the depth. Thus, detailed analysis of the minimum divergence estimator, i.e., using only \eqref{eq:beta_div2} to estimate $d$ (without additional prior information), is out of scope of this work and is left for future work.

\subsection{Pseudo-Bayesian estimation}
\label{subsec:pseudo_post}
While robust depth estimation using only point estimates is interesting, we are interested in computing measures of uncertainty about $d$ as such information can also be propagated to estimate the object depth in future frames. Thus, we adopt a Bayesian viewpoint and use the $\beta$-divergence to construct a pseudo-Bayesian method. 
Let us assume that $d$ is assigned a prior distribution $f(d)$. Note that this assumption is consistent with the method discussed in Section \ref{sec:online}, where the depth parameters of a given frame will be assigned a product of $P$ independent distributions.  

When considering the KL divergence as a similarity measure, the classical posterior distribution of $d$, can be obtained by solving 
\begin{eqnarray}
\label{eq:ELBO_KL}
 \underset{p(d) \in \mathbb{P}}{\min} L(p(d)),
\end{eqnarray}
where $\mathbb{P}$ is the set of all probability distributions, $-L(p(d))$  is the evidence
lower-bound (ELBO),
\begin{eqnarray}
\label{eq:ELBO}
L(p(d))= K \mathbb{E}_{p(d)}\left[\textrm{CE}_{\textrm{KL}}(d) \right] + D_{\textrm{KL}}(p(d)||f(d)),
\end{eqnarray}
with $\mathbb{E}_{p(d)}\left[ \cdot\right]$ the expectation with respect to $p(d)$ and where $\textrm{CE}_{\textrm{KL}}(d)=-\dfrac{1}{K}\sum_{k}\log\left(f(y_k|d) \right)$ is the cross-entropy between $\hat{f}(y)$ and $f_0(y|d)$ \cite{Futami2018}. Note that in \eqref{eq:ELBO}, the term $D_{\textrm{KL}}(p(d)||f(d))$ acts as a penalty enforcing the solution $p(d)$ to be similar to the prior distribution $f(d)$.
Indeed the solution of \eqref{eq:ELBO_KL} yields
\begin{eqnarray}
\label{eq:exact_post}
p(d)=f(d|\bfy)\propto f(d)\exp^{-K \textrm{CE}_{\textrm{KL}}(d)}.
\end{eqnarray}
As expected in \eqref{eq:exact_post}, $\exp^{-K \textrm{CE}_{\textrm{KL}}(d)}$ is indeed proportional to the likelihood $f(\bfy|d)$. 

In a similar fashion, we build a pseudo-posterior distribution, which maximize the $\beta$-ELBO $-L_{\beta}(p(d))$, i.e., 
\begin{eqnarray}
\label{eq:ELBO_beta}
 \underset{p(d) \in \mathbb{P}}{\min} L_{\beta}(p(d)),
\end{eqnarray}
where 
\begin{eqnarray}
\label{eq:L_beta}
L_{\beta}(p(d))= K \mathbb{E}_{p(d)}\left[\textrm{CE}_{\beta}(d) \right] + D_{\textrm{KL}}(p(d)||f(d)),
\end{eqnarray} and where
\begin{eqnarray}
\label{eq:beta_CE}
\textrm{CE}_{\beta}(d)=-\dfrac{\beta+1}{\beta K}\sum_{k}f(y_k|d)^\beta
\end{eqnarray}
is the $\beta$-cross-entropy between $\hat{f}(y)$ and $f_0(y|d)$ \cite{Futami2018}. 
The solution of \eqref{eq:ELBO_beta} yields
\begin{eqnarray}
\label{eq:approx_post}
p(d)=\tilde{f}(d|\bfy)\propto f(d)\exp^{-K \textrm{CE}_{\beta}(d)}.
\end{eqnarray}
The solution $\tilde{f}(d|\bfy)$ of \eqref{eq:ELBO_beta} and the traditional posterior distribution $f(d|\bfy)$ in \eqref{eq:exact_post} present very similar expressions, the main difference being the likelihood term in \eqref{eq:exact_post} which is replaced by $\exp^{-K \textrm{CE}_{\beta}(d)}$ in \eqref{eq:approx_post}. Thus, $\tilde{f}(d|\bfy)$ is referred to as pseudo-posterior distribution, as it relies on the pseudo-likelihood $\exp^{-K \textrm{CE}_{\beta}(d)}$.

While $\tilde{f}(d|\bfy)$ is generally non-standard, its mean and variance can be efficiently computed via numerical integration, e.g., by discretizing the admissible domain of definition of $d$, especially since the expected support of $d$ is bounded in practice. Thus, we use as depth point estimate the mean of the pseudo-posterior $\tilde{f}(d|\bfy)$ and as measure of the uncertainty the variance of $\tilde{f}(d|\bfy)$. In addition to providing summary statistics about the current depth, the mean and variance of the posterior distribution in \eqref{eq:approx_post} can also be incorporated in the prior model of the next frame, as will be discussed in Section \ref{sec:online}.

Although this paper focuses on robust depth estimation from single-wavelength SPL, the approach proposed here can also be used when multispectral lidar data are available. The resulting $\beta$-divergence and pseudo-posterior distribution are detailed in Appendix.

\subsection{Preliminary comparative study}
\label{subsec:prelim_results}
\begin{figure}[ht!]
	\center
	\includegraphics[width=\columnwidth]{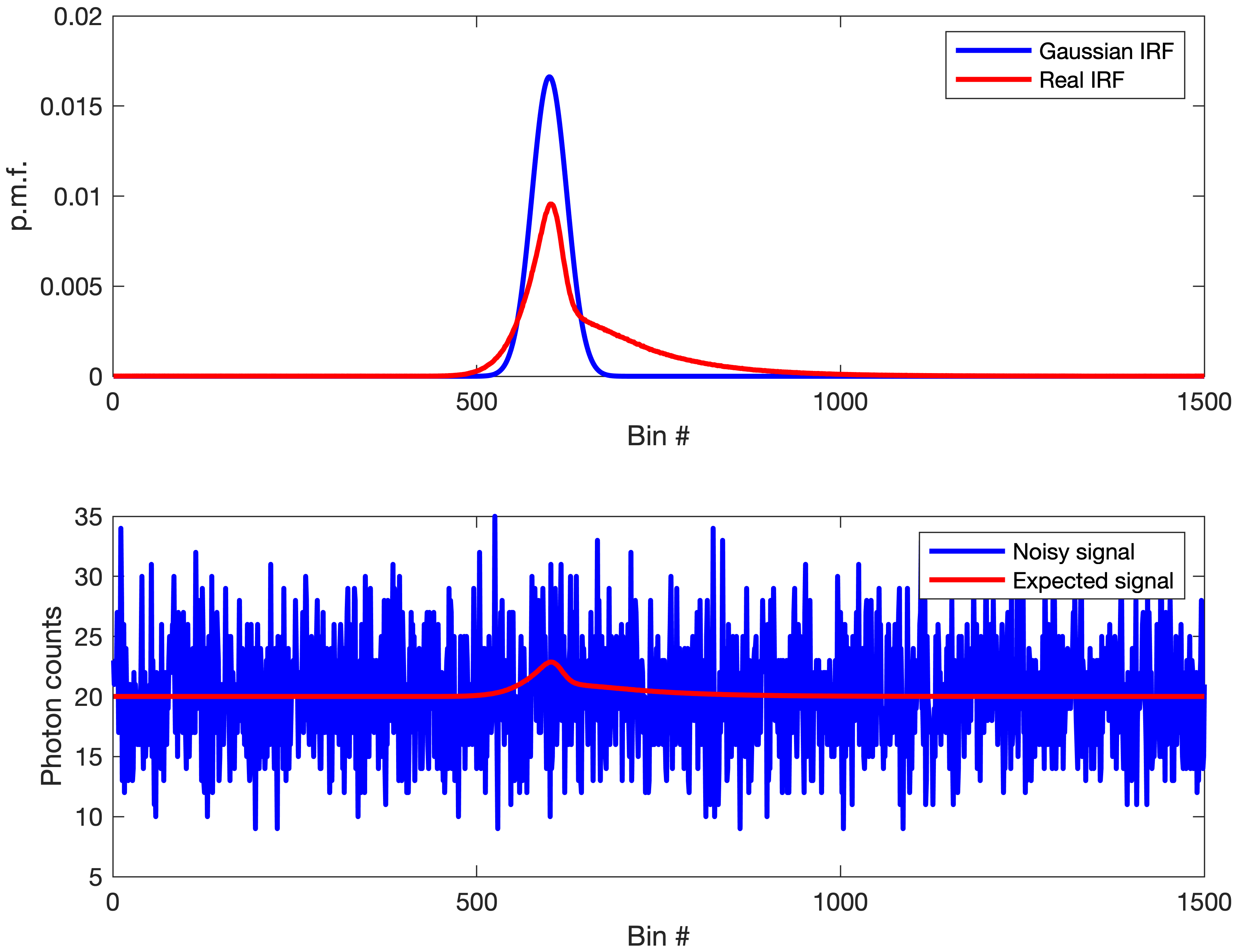}\\
	\vspace{-0.3cm}
	\caption{Top: Real (red) and Gaussian (blue) IRFs used to simulate synthetic data. The position of the peak is bin 600 for each IRF. Bottom: Example of expected (red) and noisy (blue) histogram of photon counts, using the real IRF, for $300$ signal photons and $\text{SBR}=10^{-2}$ (the peak is located at bin 600).}
	\label{fig:IRF_synth}
\end{figure}

Prior to applying the proposed robust depth estimation strategy to online 3D imaging, we assess the depth estimation quality for a single pixel. For this study, we consider two normalized IRFs with unitary integrals, depicted in Fig. \ref{fig:IRF_synth} (top). We generated synthetic data using $T=1500$ and $N_T=1500$ time bins. The first IRF (red curve) in Fig. \ref{fig:IRF_synth} is a real IRF measured in \cite{Tobin2017,Ren2018_SW-MSL} (473 nm) and presents a full width at half maximum (FWHM) of $28$ bins. Each time bin corresponds to a 2 ps time interval. The second IRF (blue curve) presents a Gaussian shape with the same FWHM. This second IRF allows us to investigate the impact of the asymmetry of the IRF on the depth estimation. For simplicity, we assume the position of the surface is associated with the time instant where the IRF has the highest amplitude. Given the high resolution of the discretization grid (compared to the shape of the IRF), using  \eqref{eq:lik_bg} or \eqref{eq:lik_bg_poisson} yields similar results thus we do not distinguish the two models. For each IRF, we generated data from \eqref{eq:lik_bg_poisson} and investigated various illumination scenarios with SBR in the interval $[10^{-4},10^{2}]$ and mean signal counts (MSC) (in each pixel) ranging from $10$ to $1000$. To illustrate the difficulty of the problem, an example of histogram with $300$ signal photons and SBR of $0.01$ is depicted in Fig. \ref{fig:IRF_synth} (bottom). For each couple of SBR/signal photon values, we generated $N_{\textrm{MC}}=2000$ histograms, drawing each depth parameter from a Gaussian distribution with mean $600$ and variance $2500$. 

Here, the depth estimation performance for a single pixel is assessed through a comparison of (pseudo-)posterior means based on a fixed depth prior distribution.
For such estimators, the prior distribution used is the Gaussian distribution with mean $600$ and variance $2500$ used to generate the data, which is a relatively weakly informative prior distribution.
The reference estimator, referred to as ``Oracle", is the minimum mean squared error (MMSE) estimator of $d$ associated with \eqref{eq:lik_bg_poisson}, assuming that $(r,b)$ is perfectly known. Similarly the MMSE estimator associated with \eqref{eq:lik_ideal_poisson} is referred to as ``BF" for \emph{background-free}. We also include a non-parametric estimator of $d$, namely the half-sample mode estimator \cite{Bickel2006}, denoted by ``HSM". Additional robust estimators could be considered, such as the Huber estimator \cite{Huber1964,Wilcox2012}. However, the latter requires sensitive parameter tuning (depending on the SBR) and does not provide satisfactory results in the low SBR regime of interest here, where the photon detections considered as outliers can represent more than $99\%$ of the detected photons. Thus, we only report the results of HSM. Finally, we consider the pseudo-posterior mean (pseudo-MMSE estimator) obtained from \eqref{eq:approx_post}. It is referred to as ``PB" for \emph{pseudo-Bayesian}.

To assess if the competing methods can accurately estimate the depth parameters, we define the empirical probability of ``satisfactory" detection \cite{TachellaEUSIPCO2019} as 
\begin{eqnarray}
p_d=\dfrac{1}{N_{\textrm{iter}}}\sum_{n=1}^{N_{\textrm{iter}}}\mathbb{I}\left(|\hat{d}_n-d_n|<\eta\right)
\end{eqnarray}
where $d_n$ (resp. $\hat{d}_n$) is the actual (resp. estimated) depth estimate and $\mathbb{I}(\cdot)$ is the indicator function, which is equal to $1$ if $|\hat{d}_n-d_n|<\eta$ and $0$ otherwise. Moreover, $\eta$ is a parameter reflecting which error is deemed acceptable. Here we set $\eta=28$ (the FWHM of the IRFs).

\begin{figure}[ht!]
	\center
	\includegraphics[width=\columnwidth]{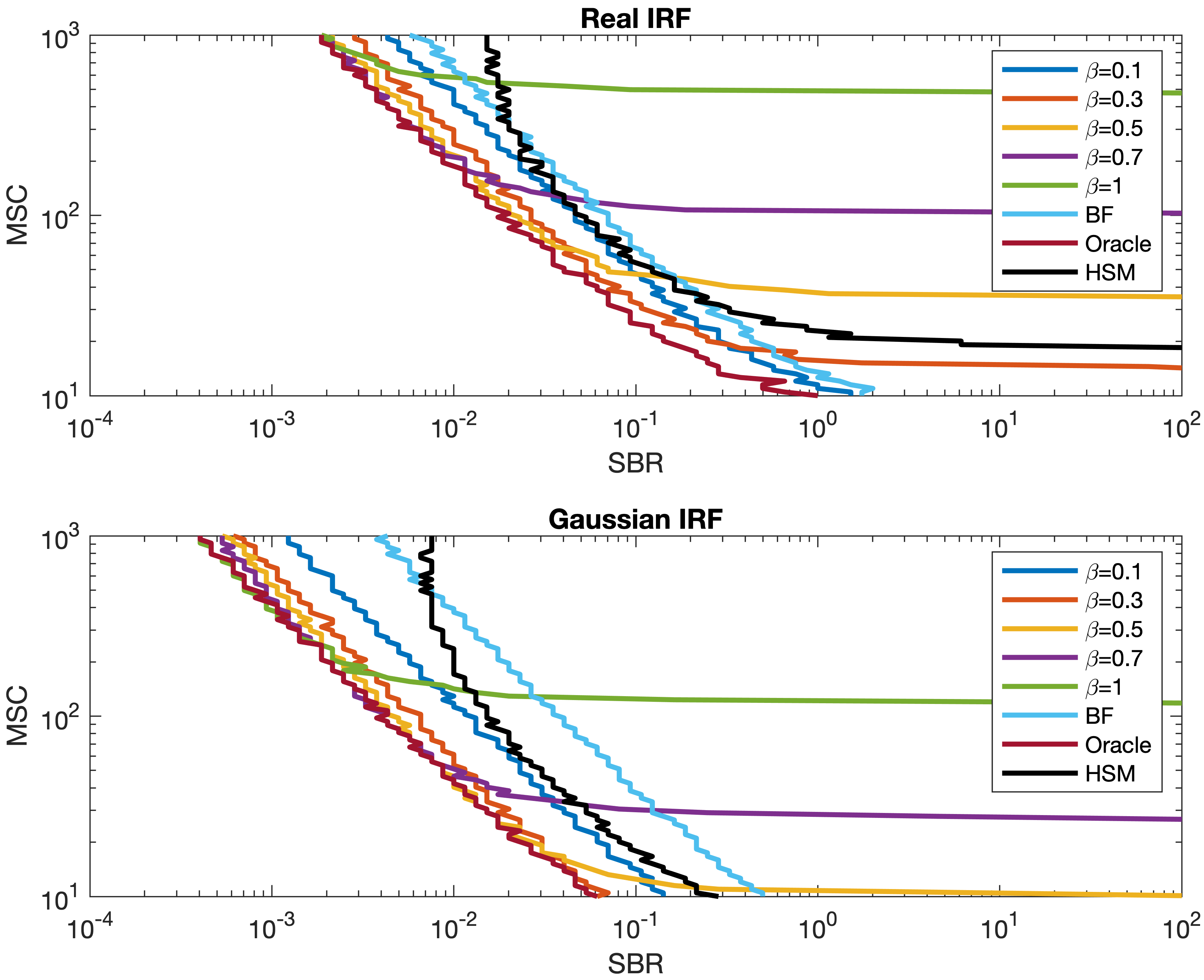}\\
	\caption{Curves of satisfactory detection ($p_d=85\%$, with $\eta=28$) as a function of the SBR and mean signal count (MSC) using the real IRF (top) and Gaussian IRF (bottom). This figure compares HSM and (pseudo-)MMSE estimators using the same Gaussian prior distribution with mean $600$ and variance $2500$.}
	\label{fig:MMSE_synth}
\end{figure}

Fig. \ref{fig:MMSE_synth} compares the depth estimation performance of HSM, Oracle, BF and PB for five values of $\beta$. This figure displays, for each method, the curve of satisfactory detection ($p_d=85\%$) as a function of the SBR and signal photon counts. On the right-hand side of each curve, each method yields $p_d>85\%$.
As mentioned above, the prior distribution has been set to be weakly informative to better highlight the behavior of the different methods. 
If the prior distribution was more informative and correct (e.g., properly centered around the actual depth value), it would dominate the (pseudo-)likelihood factors in the (pseudo-)posterior distributions and all the methods would present similar behaviors and improved performance. Fig. \ref{fig:MMSE_synth} thus illustrates how the different methods perform when limited information is available about the unknown depth. 
As expected, the Oracle provides the best results for both IRFs. The BF estimator is significantly less accurate for low SBRs. Using the PB estimators, the curves approach those of BF for small values of $\beta$, while they tend to converge towards the Oracle curves when increasing $\beta$, provided that the MSC is large enough. This figure also shows that performance of all the estimators depends on the skewness of the IRFs.

Overall, HSM leads to less accurate results than the other methods, partly because it does not leverage the shape of the underlying IRF, and because it does not use additional prior information. Note also that HSM is not accurate when the MSC is low since the mode of the distribution then becomes difficult to estimate (see Fig. \ref{fig:MMSE_synth} (top)).

Irrespective of the IRF shape, the parameter $\beta$ of the divergence plays a key role in the working region of the resulting algorithm. For a given MSC level, increasing $\beta$ allows a reduction of the limiting SBR below which the algorithm starts to fail. This limiting SBR however remains bounded by the limiting SBR of the Oracle. On the other hand, increasing $\beta$ also increases the minimum MSC required for the algorithm to perform satisfactorily. For instance, in Fig. \ref{fig:MMSE_synth} (top), with $\beta=0.5$ it is still possible to estimate the depth accurately for $\textrm{SBR}>1$ and $35$ signal photons while setting $\beta=0.7$ in such scenarios leads to poor results, potentially worse than BF. Setting $\beta$ depends on shape of the IRF but also on the expected SBR/MSC. Note that if the SBR is sufficiently large, the background effect is not significant and the depth reconstruction does not require a robust method. 
These first results show near-optimal results can be obtained with PB for large values of $\beta$ (provided that MSC is large enough), without estimating additional model parameters and with a fixed computational budget.

\begin{figure}[ht!]
	\center
	\includegraphics[width=\columnwidth]{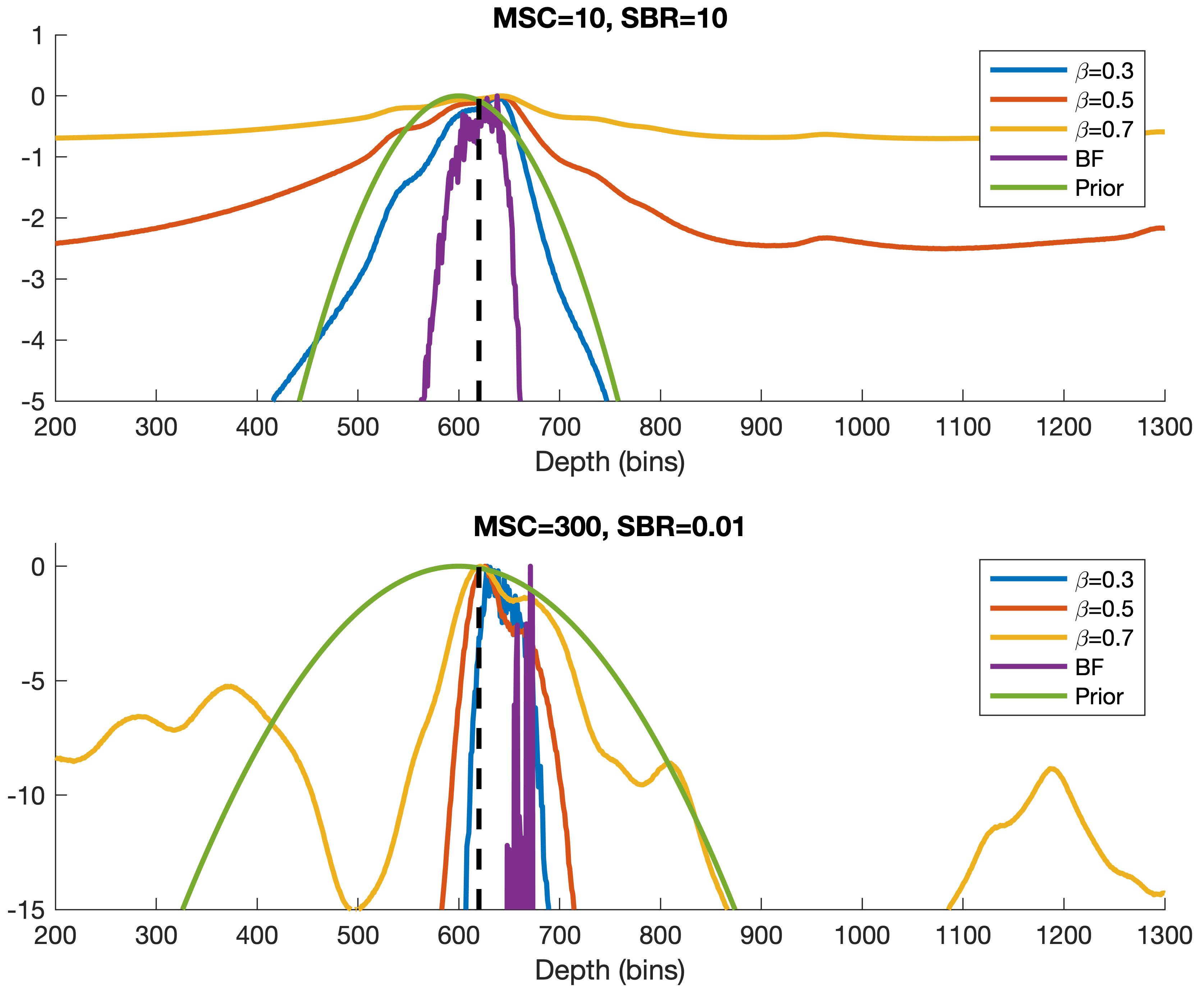}\\
	\caption{Comparison of the log-likelihood (BF) and log-pseudo-likelihood terms (PB) for (MSC, SBR) $=(10,10)$ (top) and (MSC, SBR) $=(300,0.01)$ (bottom), as a function of $d \in [200,500]$. The dashed black lines represent the actual position of the object ($d=620$). Note that the scale of the y-axis is different in the two plots.}
			\vspace{-0.3cm}
	\label{fig:compare_beta}
\end{figure}

To better illustrate the impact of the $\beta$-divergence in the various MSC/SBR scenarios, we show in Fig. \ref{fig:compare_beta} how the parameter $\beta$ affects the pseudo likelihood term and in turn the pseudo-posterior distribution. In these two examples, we generated data with (MSC, SBR) $=(10,10)$ and (MSC, SBR) $=(300,0.01)$. The actual position of the object is at bin 620 and the Gaussian prior distribution has a mean of $600$ and variance $2500$. The curves in Fig. \ref{fig:compare_beta} represent the logarithm of the (pseudo)-likelihood terms for different values of $\beta$, the likelihood term assuming no background (BF) and the log-prior. All the curves are normalized such that the maximum of each curve is 0 since vertical offsets do not affect the posterior distribution.

When (MSC, SBR) $=(10,10)$ (Fig. \ref{fig:compare_beta}, top plot), the likelihood \eqref{eq:lik_ideal} is similar to the actual distribution of the data and the maximum of the log-likelihood is located around the actual depth value. When $\beta$ increases, the pseudo-likelihood term becomes flatter, which gives more weight to the prior distribution in the pseudo-posterior distribution in \eqref{eq:approx_post}. Thus the mean of the pseudo-posterior distribution tends to the mean of the prior distribution. In this regime, small values of $\beta$ should be preferred. When (MSC, SBR) $=(300,0.01)$ (Fig. \ref{fig:compare_beta}, bottom plot), the likelihood \eqref{eq:lik_ideal} becomes very concentrated, potentially around a value that is far from the actual depth due to the mismatch between the background-free observation model and the actual distribution of the data. This likelihood term is more concentrated than the prior and thus dominates the posterior distribution, leading to a poor MMSE depth estimate. In a similar fashion to the previous scenario, when $\beta$ increases, the pseudo-likelihood term becomes flatter, which gives more weight to the prior distribution in the pseudo-posterior distribution in \eqref{eq:approx_post}. Note that the mode of the pseudo-likelihood can also change when $\beta$ changes. For instance, the mode for $\beta=0.3$ is much closer to the actual depth than when using the likelihood \eqref{eq:lik_ideal}. As discussed when analyzing Fig. \ref{fig:MMSE_synth}, in this regime, larger values of $\beta$ are preferred.

For practical applications, it is thus important to select $\beta$ sensibly based on the expected observation conditions, balancing performance at low MSCs and low SBRs. Fortunately, it is possible to pre-compute performance bounds as in Fig. \ref{fig:MMSE_synth}, for any real instrumental response. Moreover, from the preliminary experiments we conducted, it seems that using $\beta \in [0.3,0.6]$ leads to a good trade-off for both the low SBR and low MSC regimes.

\subsection{Target detection and additional parameter estimation}
\label{subsec:detection}
In Section \ref{subsec:pseudo_post}, we proposed to estimate the depth of a surface in a given pixel based on \eqref{eq:lik_ideal}.
This model assumes that an object is actually present in the pixel considered, which is not always true, especially for long range imaging applications. Thus, it is important to be able to decide whether a surface is actually present and this cannot be achieved directly using \eqref{eq:approx_post}. To address this problem, we use the Bayesian object detection algorithm proposed in \cite{TachellaSPIE2019}. This method uses the Poisson likelihood model in \eqref{eq:lik_bg_poisson} and assigns prior distributions $f(r), f(b)$ and $f(d)$, to the reflectivity $r$ of the target, the background level $b$ and the target depth, respectively. The detection is finally seen as a binary hypothesis test where the background, and the target reflectivity and depth are marginalized. More precisely, the algorithm decides, a posteriori, whether $r=0$ (no surface) or $r>0$ (surface present). The output of the algorithm is $\pi$, the posterior probability of target presence, which can then be thresholded to derive a detection map (the interested reader is invited to consult \cite{TachellaSPIE2019,TachellaEUSIPCO2019} for additional details about the detection method). Note that the algorithm also incorporates the prior probability of target presence $\pi^0$ (through a spike-and-slab prior model $f(r)$). This point will be further discussed in Section \ref{subsec:online_detection}, together with the choice of the other prior distributions used for object detection. 

After the pixel-wise detection procedure, it is possible (if needed) to estimate the average background level and the target reflectivity (if a target is detected), for instance using \eqref{eq:lik_bg_poisson}. To keep the computational overhead low, we only report a fast method here, but more complex schemes, as in \cite{shin2015photon,altmann2016lidar,Pawlikowska:17} could be used. If no target is detected, the background level is estimated by dividing the photon count $K$ by $T$. If an object is detected, we use as depth point estimate the mean of the pseudo-posterior distribution in \eqref{eq:approx_post}, and the target reflectivity is estimated together with the background by MLE using \eqref{eq:lik_bg_poisson}.

\section{Application to online reconstruction}
\label{sec:online}
In this section, we consider a set of $N$ sequential temporal periods during which a ToA histogram is recorded for each of the $P$ pixels. We denote by $\bfy_{n,p}$ the set of photon ToAs and $d_{p,n}$ the depth of the object in the pixel $p$ and frame $n$.

\subsection{Approximation using Assumed Density Filtering}
As mentioned in the introduction, our online estimation procedure consists of leveraging the temporal correlation between successive frames by incorporating the posterior distribution of the depth profile at time $(n-1)$ in the inference problem at time $n$. As described in \cite{Altmann2019}, estimating the posterior mean and variance of $d_{p,n}$ presents a significant advantage beyond simply providing summary statistics about the current range profile. It allows the derivation of tractable adaptive estimation procedures. A classical choice for modeling relatively slowly evolving parameters is the Gaussian random walk (RW), i.e., 
\begin{eqnarray}
f_t(d_{p,n}|d_{p,n-1}) \propto \exp\left\lbrace-\dfrac{(d_{p,n}-d_{p,n-1})^2}{2\sigma_{\textrm{RW}}^2}\right\rbrace,
\end{eqnarray}
controlled by the variance $\sigma_{\textrm{RW}}^2$. This RW mostly allows displacements smaller than $3\sigma_{\textrm{RW}}$ along the direction of the observation (using the 3-sigma rule of thumb). Whilst this approach is simple, it does not allow for rapid changes as might occur when the imaging system or the scene moves orthogonally to the direction of observation. To alleviate issues associated with such changes while keeping the estimation strategy tractable, we define as in \cite{Altmann2019}, for each pixel, a local neighborhood $\mV_p$ of $M$ neighbors (including the current pixel) and define the following prior model\\
$f(d_{p,n})$
\begin{eqnarray}
\label{eq:predict}
\propto \sum_{p' \in \mV_p} \nu_{p'}\int f_{t}(d_{p,n}|d_{p',n-1})q_{p',n-1}(d_{p',n-1})\textrm{d} d_{p',n-1},
\end{eqnarray}
where $\{q_{p,n-1}(\cdot)\}_{p}$ are Gaussian distributions. Basically, the prior model of $d_{p,n}$ is constructed via a Gaussian mixture model using the depth information in neighboring pixels at the previous frame, convolved by a Gaussian RW. In \cite{Altmann2019}, $\{q_{p,n-1}(\cdot)\}_{p}$ was the set of Gaussian approximations of the depth posterior distributions obtained at frame $(n-1)$. A similar approach is adopted here since the Gaussian approximations of the posterior distributions can be obtained as for assumed density filtering (ADF) \cite{Lauritzen1992,Boyen1998} and expectation-propagation\cite{minka2001expectation}, i.e., by minimizing the KL divergence
\begin{eqnarray}
\label{eq:KL_div}
D_{\textrm{KL}} \left[\tilde{f}(d_{p,n-1}| \bfy_{p,n-1}) || q_{p,n-1}(d_{p,n-1})\right]
\end{eqnarray} 
w.r.t. $q_{p,n-1}(d_{p,n-1})$ which belongs to the family of Gaussian distributions and where $\tilde{f}(d_{p,n-1}| \bfy_{p,n-1})$ is the pseudo-posterior distribution described in Section \ref{subsec:pseudo_post}. This minimization reduces to matching the mean and variance of $\tilde{f}(d_{p,n-1}| \bfy_{p,n-1})$ and $q_{p,n-1}(d_{p,n-1})$, hence the discussion about the estimation of the moments of $\tilde{f}(d_{p,n-1}| \bfy_{p,n-1})$ in Section \ref{subsec:pseudo_post}.

However, one of the main limitations of the spatio-temporal model used in \cite{Altmann2019} is that it does not explicitly take into account whether objects were actually present in the neighboring pixels in frame $(n-1)$ when building the prior model for frame $n$. To address this problem, we incorporate the results of the detection procedure detailed in Section \ref{subsec:detection}. If an object is detected in pixel $p'$ and frame $(n-1)$, $q_{p',n-1}(\cdot)$ is set to the Gaussian approximation of the pseudo-posterior \eqref{eq:approx_post} in that pixel, as in \cite{Altmann2019}. If no object is detected in pixel $p'$ and frame $(n-1)$, $q_{p',n-1}(\cdot)$ is replaced by a Gaussian distribution with mean $(d_{\textrm{min}}+d_{\textrm{max}})/2$ and variance $(d_{\textrm{max}}-d_{\textrm{min}})^2/12$, where $(d_{\textrm{min}},d_{\textrm{max}})$ are the expected lower and upper bound of the scene depth. This choice of mean and variance leads to a flat prior distribution mimicking the uniform distribution defined on $(d_{\textrm{min}};d_{\textrm{max}})$. Note that in practice, $T$ is chosen large enough so that for any depth in $(d_{\textrm{min}};d_{\textrm{max}})$, the shape of $s_0(y-2d/c)$ remains the same. 
Increasing the variance of the Gaussian distributions $q_{p',n-1}(\cdot)$ of empty pixels allows us to better detect new objects appearing in the scene and random depths. A similar strategy is adopted at the edges of the image where pixels keep $M$ neighbors, some of them being outside the field of view and contributing to the mixture with weakly informative Gaussian distributions.

The $M$ weights of the mixture in \eqref{eq:predict} are set to 
\begin{eqnarray}
\nu_{p'}= \left\{
    \begin{array}{ll}
        \nu_0 \in [0,1],  & \mbox{if } p'=p \\
        \dfrac{(1-\nu_0)}{M-1} & \mbox{otherwise}, 
    \end{array}
\right.
\end{eqnarray}
where $\nu_0$ is a user-defined weight which controls the weight assigned to the central pixel of each neighborhood.  

\subsection{Online target detection}
\label{subsec:online_detection}

\begin{algogo}{R3DSP algorithm}
\label{algo:algo1}
 \begin{algorithmic}[1]
			\STATE \underline{Fixed input parameters:} Variance of RW for dynamic model: $s^2$, Neighborhood size $M$, parameter of GMM $\nu_0$, indices of faulty pixels.
			\STATE \underline{Initialization ($n=0$)}
    			\STATE Set $(q_{p,0}(\cdot), \pi_{p,0}^0), \forall p$. 
			\FOR{$n=1,\ldots N$}
				\FOR{$p=1,\ldots P$}
    			\STATE Compute the prior model $f(d_{p,n}), \forall p$ from \eqref{eq:predict} via ADF. 
				\STATE Compute the pseudo-posterior $f(d_{p,n}| \bfy_{p,n})$ in \eqref{eq:approx_post}.
				\STATE Compute $q_{p,n}(d_{p,n})$ using \eqref{eq:KL_div}.
				\STATE Estimate $\pi_{p,n}$ using \cite{TachellaSPIE2019}.
				\IF{target present, i.e., $\pi_{p,n}>0.5$}
					\STATE Set the estimated depth $\hat{d}_{p,n}$ as the mean of $q_{p,n}(d_{p,n})$.
					\STATE Estimate current background level and target reflectivity.
				\ELSE
					\STATE Set $\hat{d}_{p,n}=\emptyset$.
					\STATE Estimate current background level.
				\ENDIF
				\ENDFOR	
				\STATE Compute $\{\pi_{p,n+1}^0\}, \forall p$ using \eqref{eq:prior_detect}.
			\ENDFOR
\end{algorithmic}
\end{algogo}

As mentioned in Section \ref{subsec:detection}, the target detection algorithm proposed in \cite{TachellaSPIE2019} requires, for each pixel of the frame $n$, prior distributions for the background level, the target depth, its reflectivity and a prior probability of target presence to be refined. An exponential prior model is chosen for the background, whose mean is given by the background estimate obtained at that pixel in the previous frame, assuming that the background varies slowly over time. Similarly, let $\{\pi_{p,n-1}\}_p$ be the probabilities of target presence estimated at frame $(n-1)$. The prior probabilities of target presence $\{\pi_{p,n}^0\}_p$ of the frame $n$ are computed by nonlinear local averaging 
\begin{eqnarray}
\label{eq:prior_detect}
\pi_{p,n}^0 = \sigma\left(\sum_{p' \in \mV_p} \nu_{p'} \sigma^{-1}\left(\pi_{p,n-1} \right)\right),
\end{eqnarray}
where $\sigma(x)=(1+\exp(-x))^{-1}$ is the logistic function. Nonlinear averaging is preferred here over simple averaging as it promotes values closer to $0.5$, and is hence less informative. As in \cite{TachellaSPIE2019}, the prior model for the reflectivity (assuming a target is present) is a gamma distribution whose parameters are set using the calibration measurements (which provide insight about the expected signal photon counts when an object is present). In contrast to the background prior, the depth prior model used for the detection step is not set using a predictive model such as \eqref{eq:predict}. Instead, it is set in an empirical Bayes fashion using the pseudo-posterior computed in \eqref{eq:approx_post}.

The pseudo-code of the proposed method, referred to as R3DSP (for Robust 3D reconstruction using Single-Photon data) is presented in Algo. \ref{algo:algo1}. While the pseudo-code includes a loop of the $P$ pixels for each frame, it is important to remark that all the pixels of a frame can be processed independently and in parallel fashion. Thus the resulting algorithm is scalable and well adapted for GPU-based implementation. Note that the proposed algorithm can also be applied in the presence of faulty pixels for which no observations are available. In that case, we simply set  $\tilde{f}(d_{n,p}|\bfy_{np})=f(d_{p,n})$ and $\pi_{p,n}=0.5, \forall n$.

\section{Results}
\label{sec:results}
To demonstrate the benefits of the proposed algorithm, we used a bistatic transceiver system incorporating a Princeton Lightwave Kestrel camera that was capable of providing picosecond resolution, time-tagged, single-photon data from its $32 \times 32$ SPAD detector array, which captures 150,400 binary frames per second (see \cite{Tachella2019_RT3D} for more details). The transceiver system used a sub-nanosecond pulsed laser source operating at a wavelength of $1550$ nm to flood-illuminate the scene of interest.  We acquired a series of 3D videos using $T=153$ histogram bins (binning resolution of $3.75$ cm) and here, we report results obtained from two videos, both measured in daylight conditions with significant ambient light background. 

In the first experiment, we integrated the binary acquisitions into 500 lidar frames per second. At this frame rate, each lidar frame is composed of about 300 binary frames, i.e., contains at most 300 photons (see \cite{Tachella2019_RT3D} for additional details about the experimental setup). We considered a dynamic scene which consists of two people, standing approximately 1.5 metres apart, exchanging a $\approx 220$ mm diameter ball at a distance of ~320 metres from the lidar system. 
Note that in this configuration, the depth estimation performance is not expected to be altered by the distance of the observed objects within the range associated with the detector gate (4-5 meters here).
Approximately half of the pixels do not contain any surface and a single peak is usually observed in the remaining pixels (either one of the two pedestrians or the ball). In each pixel and frame, we observe approximately 35 photons related to dark counts or ambient illumination from solar background, and the visible surfaces lead to 55 additional photons per pixel, on average. For those pixels, the SBR is thus $\approx 1.6$. The IRF of each pixel was recorded during the system calibration. Although they could have been approximated by Gaussian IRFs, we used the actual IRFs during our analysis as they are slightly skewed. The proposed algorithm has been applied to a series of $3000$ successive frames, representing a total acquisition of $6$ s. For this scenario, we set, $M=5$ neighbors, $\sigma_{\textrm{RW}}=\sqrt{3}$ bins, $\nu_0=0.5$ and $\beta=0.5$. However, we did not notice significant changes when using $\beta \in [0.4,0.7]$. The reconstructed point clouds, together with a standard video of the scene recorded by a camera located next to the two people are presented in the \href{https://youtu.be/nSoLT0FbMmU}{Video 1}. In all the videos presented in this work, the colormap of the point cloud represents the amplitude (number of signal photons) of the returns in the lidar data. 

We compared the performance of the proposed method to that of two methods able to handle rapidly thousand of frames. First, we consider the classical depth MLE estimator BF assuming \eqref{eq:lik_ideal_poisson} applied independently to each pixel, and followed by an intensity-based thresholding step (5$\%$ of the IRF intensity) using the intensity estimated via MLE and \eqref{eq:lik_bg}. This thresholding step allows us to remove estimated surfaces which present too low an intensity. The second method is the RT3D algorithm \cite{Tachella2019_RT3D} recently proposed for fast reconstruction of complex (multi-surface) scenes. While RT3D aims at solving a more complex problem, that is, the estimation of an unknown number of peaks per pixel, it possesses a surface detection capability which is of interest in our study. The parameters of RT3D have been tuned via cross-validation by optimizing the visual quality of the reconstruction. In particular, although the data consists only of $32 \times 32$ pixels, RT3D is set so that it reconstructs point clouds with $96 \times 96$ pixels in the transverse direction. Note that we also applied the pre-trained implementation (provided by the authors) of the deep-learning method proposed in \cite{Lindell:2018:3D}. The results obtained with this first dataset are similar to those obtained by other competing methods. However, this method performed worse than the other methods when considering noisier measurements (discussed next). For this reason, we did not include this method for comparison in this study. However, these first results could potentially be improved by retraining the original network in future work.

\begin{figure}[ht!]
\centering
   \begin{minipage}[b]{.49\linewidth}
     \centering
	\includegraphics[trim={2cm 0cm 2cm 1.2cm},clip,width=0.99\linewidth]{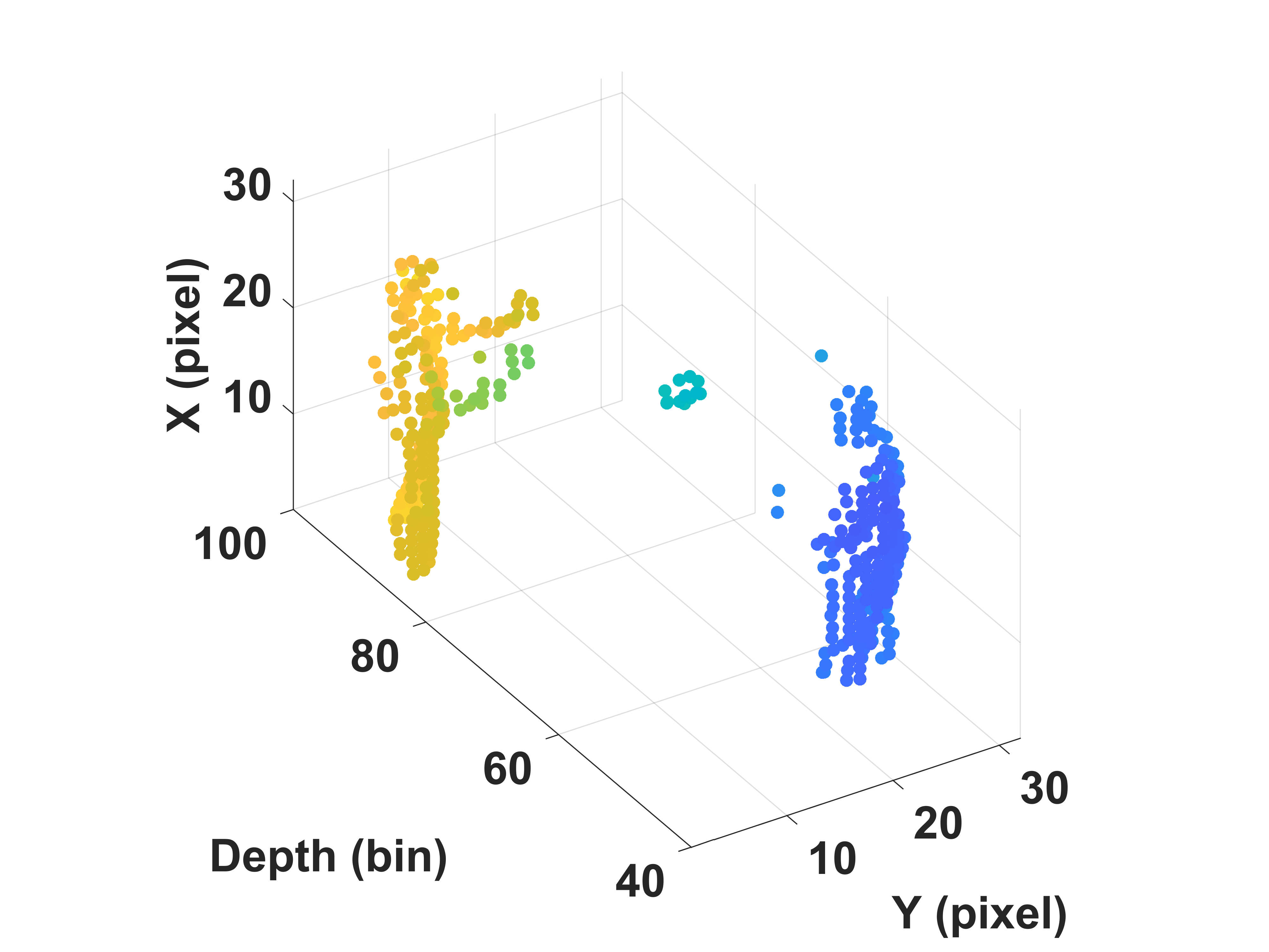}
     \subcaption{}\label{1 a}
   \end{minipage}
   \hfill
      \begin{minipage}[b]{.49\linewidth}
     \centering
	\includegraphics[trim={2cm 0cm 2cm 1.2cm},clip,width=0.99\linewidth]{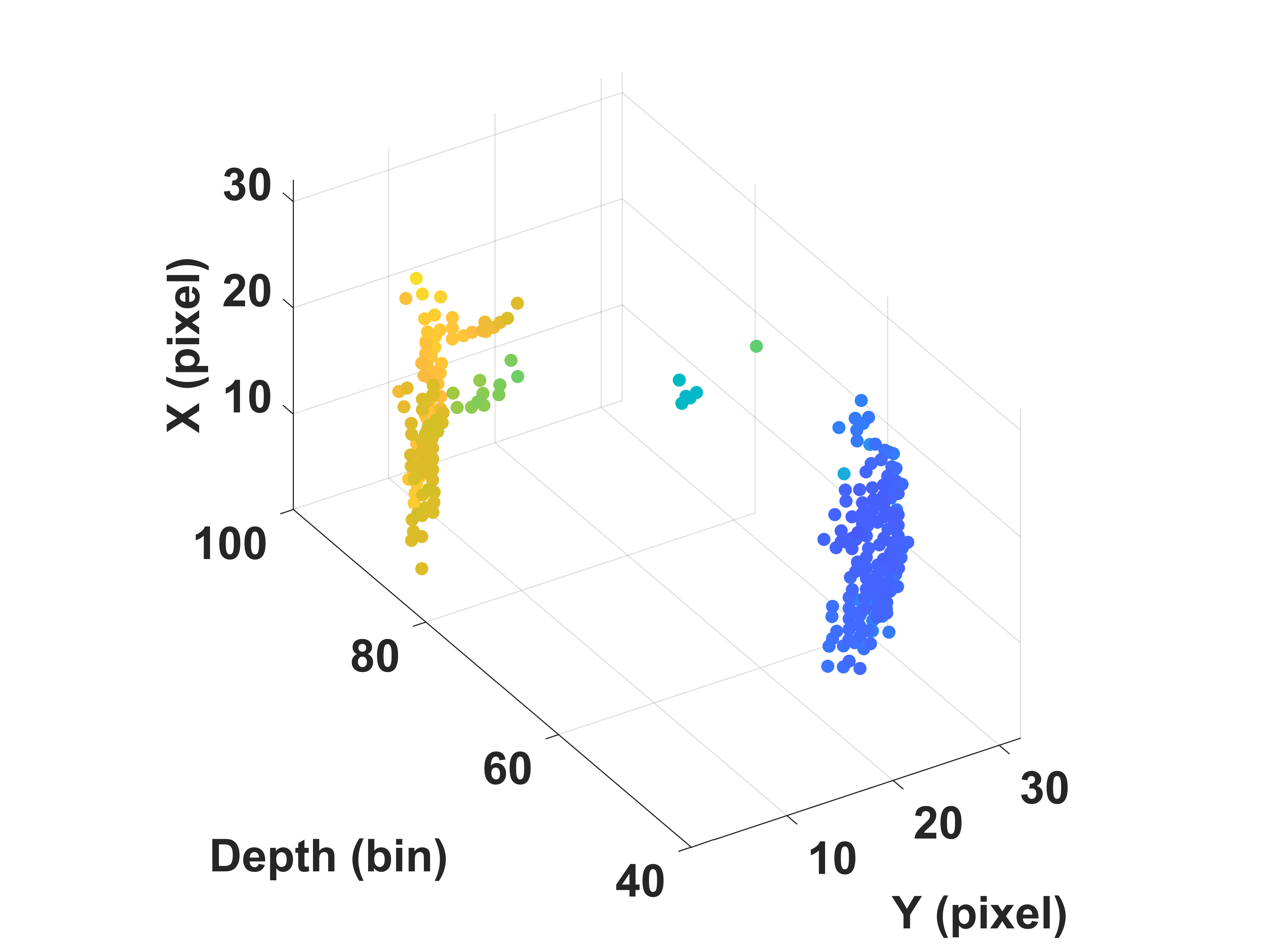}
     \subcaption{}\label{1 b}
   \end{minipage}   \\
   \begin{minipage}[b]{.49\linewidth}
     \centering
	\includegraphics[trim={2cm 0cm 2cm 1.2cm},clip,width=0.99\linewidth]{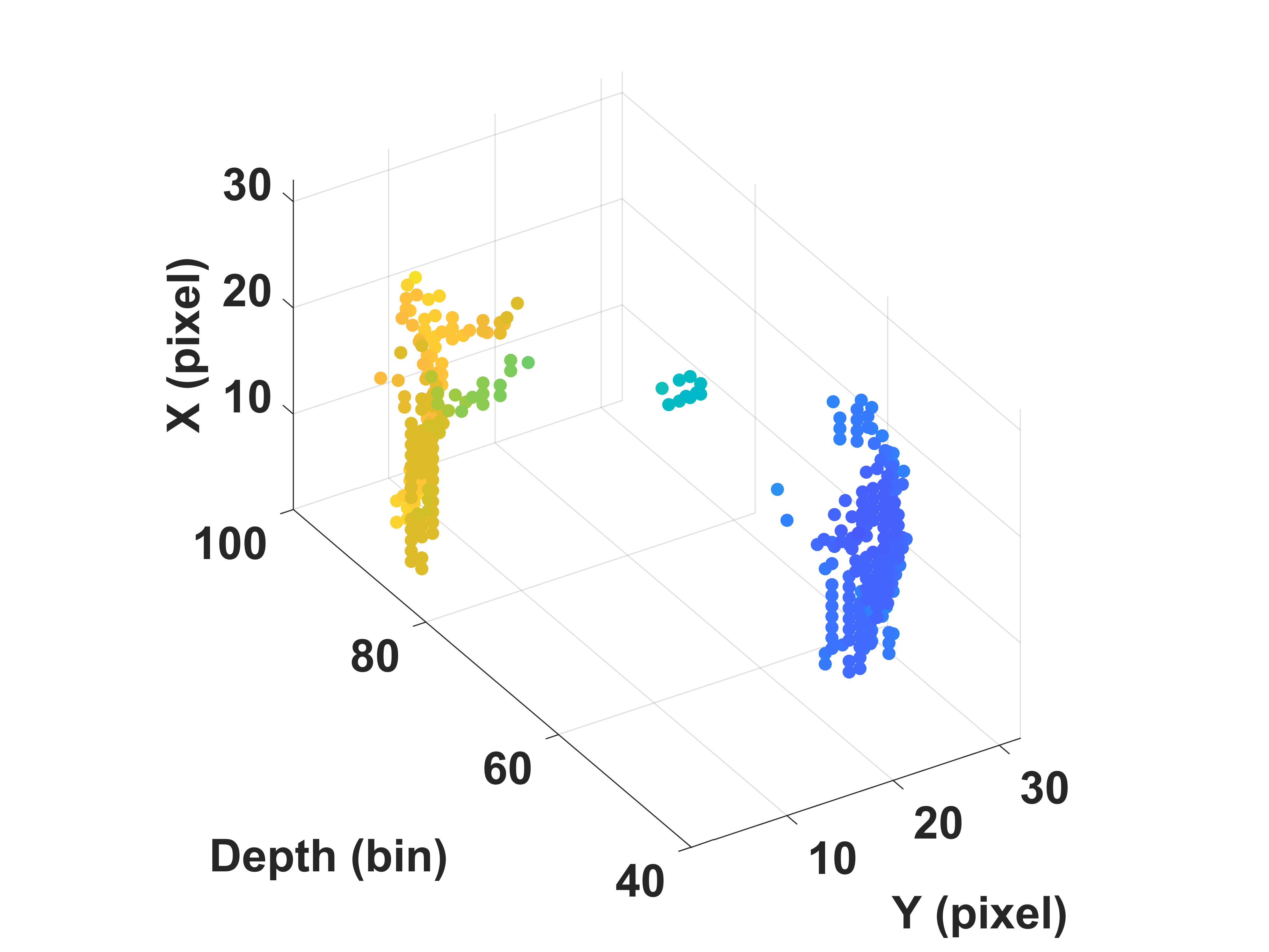}
     \subcaption{}\label{1 c}
   \end{minipage}
   \hfill
      \begin{minipage}[b]{.49\linewidth}
     \centering
	\includegraphics[trim={2cm 0cm 2cm 1.2cm},clip,width=0.99\linewidth]{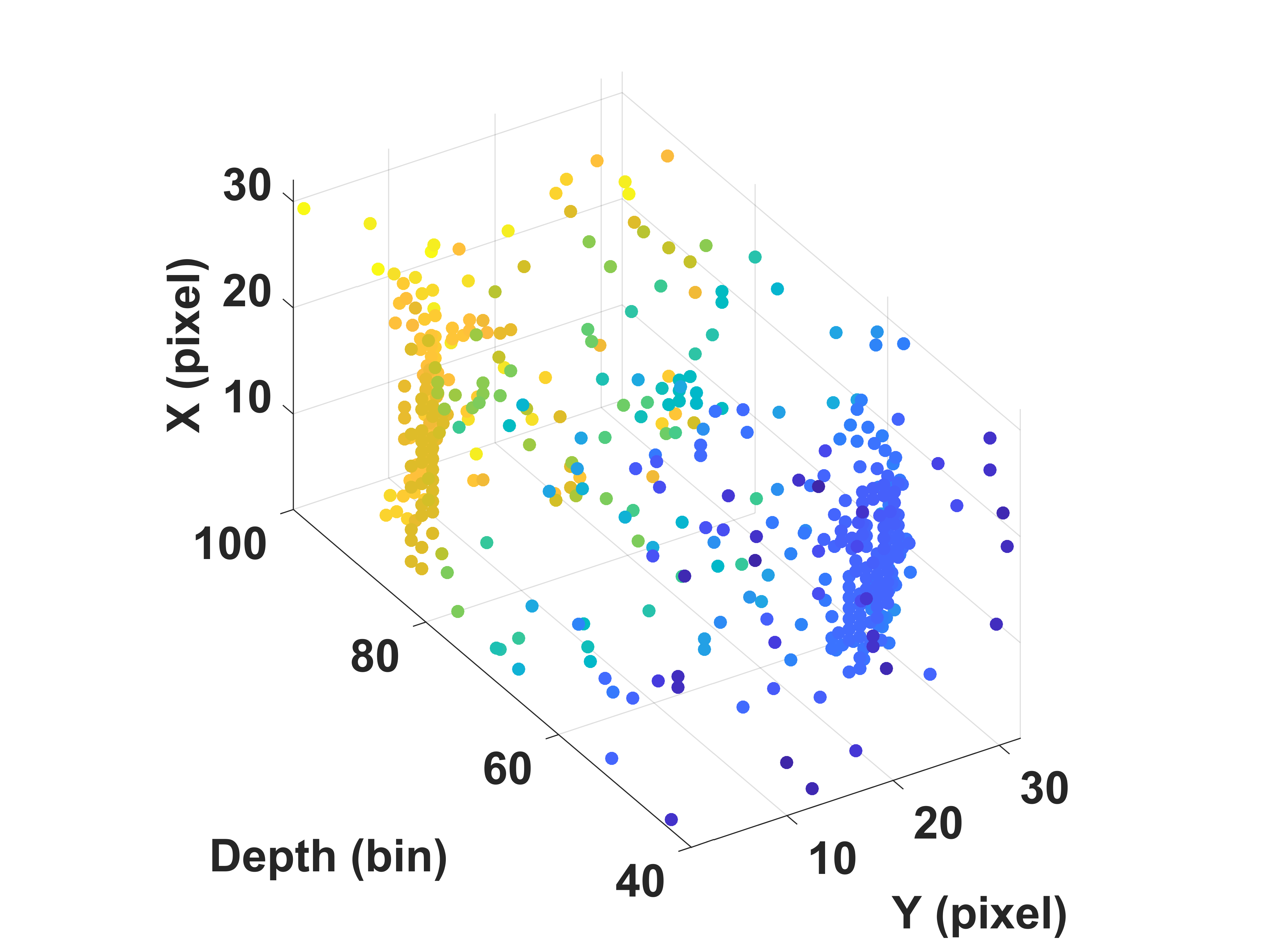}
     \subcaption{}\label{1 d}
   \end{minipage}
    \\
   \begin{minipage}[b]{.49\linewidth}
     \centering
	\includegraphics[trim={2cm 0cm 2cm 1.2cm},clip,width=0.99\linewidth]{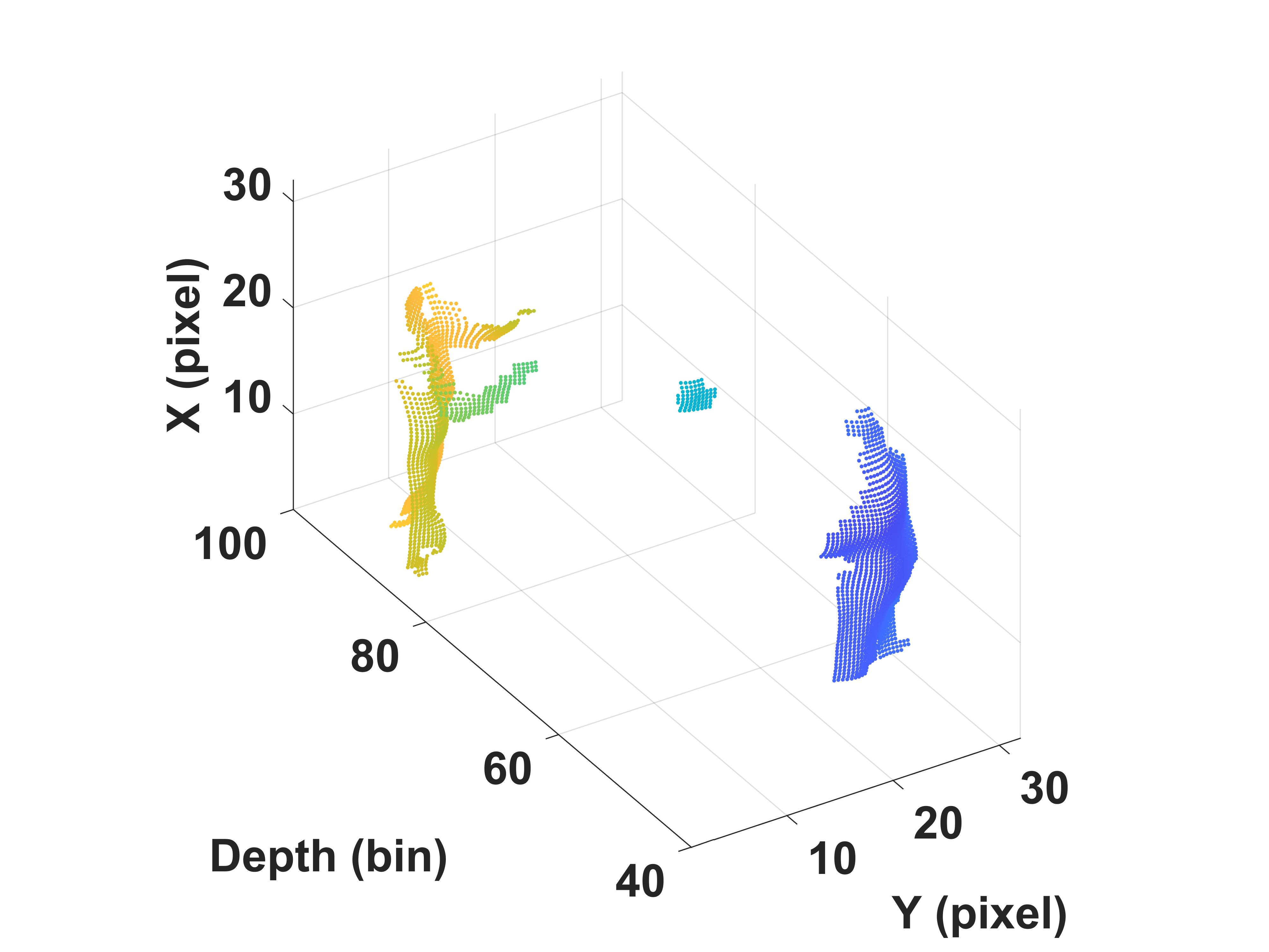}
     \subcaption{}\label{1 c}
   \end{minipage}
   \hfill
      \begin{minipage}[b]{.49\linewidth}
     \centering
	\includegraphics[trim={2cm 0cm 2cm 1.2cm},clip,width=0.99\linewidth]{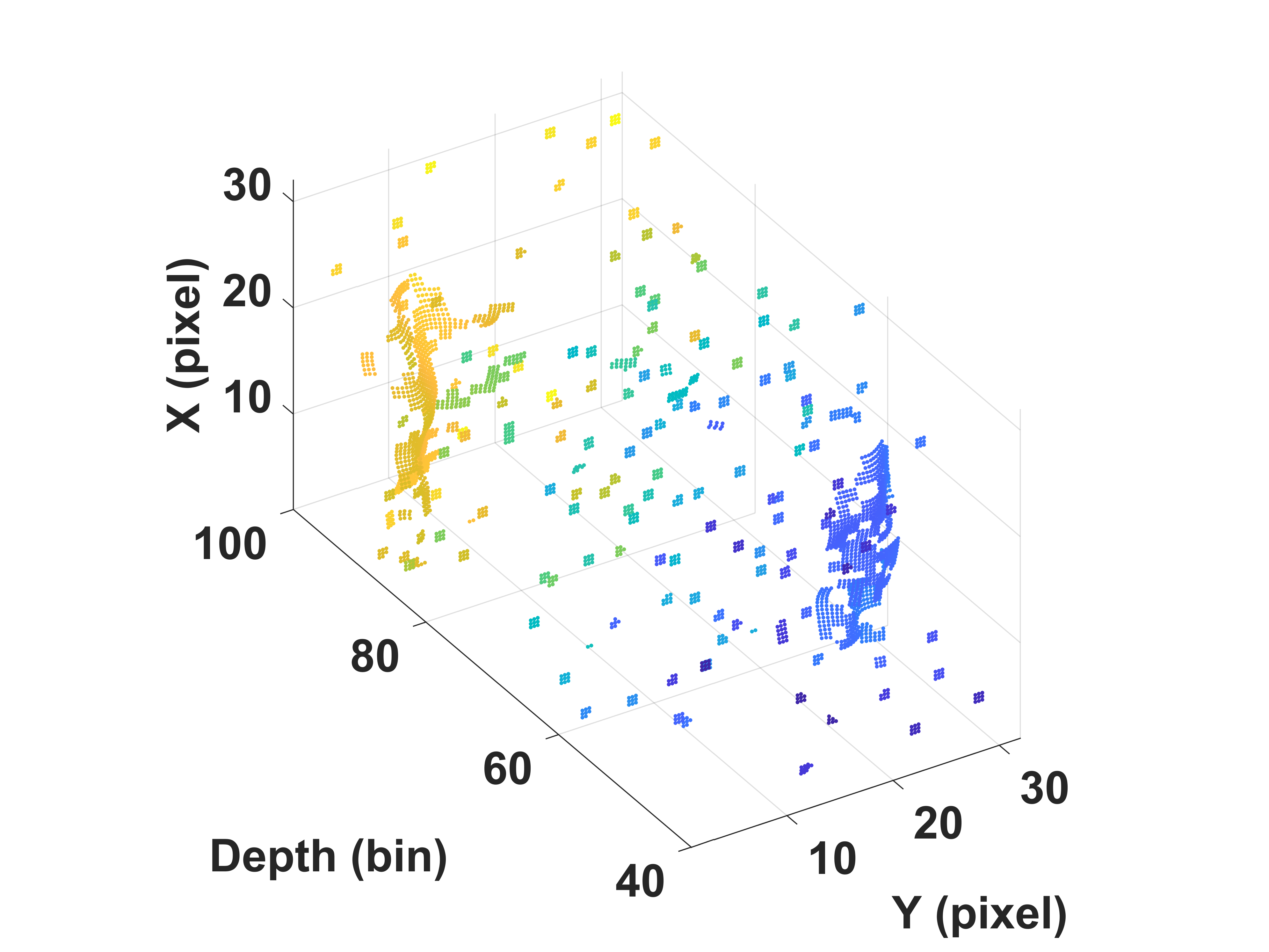}
     \subcaption{}\label{1 d}
   \end{minipage}
\caption{Examples of 3D reconstruction using the proposed method ((a),(b)), BF ((c),(d)) and RT3D (e),(f). The left-hand plots correspond to actual measurements while the right-hand plots have been obtained from the pseudo-synthetic data generated with additional background ($b=50$). The colormap represents the surface depth (axial direction).}
\label{fig:compare_ball} 
\end{figure}

Fig. \ref{fig:compare_ball} (a), (c) and (e) depict an example of reconstruction (Frame \#1700) using R3DSP, BF and RT3D, respectively. In this frame, the two pedestrians are in the field on view and the ball is roughly at the midpoint between them.  In this scenario, the three methods provide similar results, as the SBR and MSC are high enough to allow a clear identification of the empty pixels and a satisfactory estimation of the surface depth in the other pixels. Note that the surfaces appear smoother using RT3D, partly because of the spatial smoothing involved but also because the corresponding point cloud has about 9 times more points than those of R3DSP and BF. For completeness, the point clouds and background levels estimated by the three methods for the whole sequence are presented in  \href{https://youtu.be/U7bmB91XWzA}{Video 2}. For visualization purposes, the video is played at actual speed, with 50 frames per second (the intermediate frames processed by R3DSP are not displayed). This video also compares the estimated background levels, which are consistent across the three methods and it presents the estimated surface presence maps. The proposed method is able to more efficiently detect the head of the pedestrian on the left-hand side, which presents a low reflectivity due to the wavelength used ($1550$ nm). 

These results are used as reference to investigate more challenging scenarios, with lower SBRs. More precisely, we generated additional pseudo-synthetic datasets by artificially adding constant background levels to all the pixels of each sequence. We created two sequences using background levels of $b=20$ and $b=50$, leading to approximately $3060$ and $7650$ additional background photons per pixel and per frame. The resulting SBRs are $1.8\times10^{-2}$ and $7.2\times10^{-3}$, respectively. Examples of reconstructed point clouds in the lowest SBR regime ($b=50$) are depicted in Fig. \ref{fig:compare_ball} (b), (d) and (f). When the SBR decreases, BF and RT3D, which are based on intensity thresholding, generally present higher false alarm rates than R3DSP which incorporates a Bayesian test for object detection. If the thresholds of BF and RT3D are increased, the corresponding probabilities of detection decrease. Although R3DSP misses some surfaces (see Fig. \ref{fig:compare_ball} (a) and (b)), it is able to reconstruct the ball, using both the tailored detection strategy and its ability to promote correlation between successive frames via the spatio-temporal model. For completeness, the sequences reconstructed by the three methods for $b=20$ and $b=50$ are presented in \href{https://youtu.be/WB7Je8tBeDE}{Video 3} and \href{https://youtu.be/DoGgUiHRzYU}{Video 4}, respectively. These videos shows that although the three methods provide similar background estimates, R3DSP consistently yields better reconstructions (visually lower false alarm rates and higher probabilities of detection). These results confirm that in the presence of significant background levels (low SBR) the proposed method is able to detect and track dynamic surfaces more efficiently.

\begin{figure}[ht!]
	\center
	\includegraphics[width=\columnwidth]{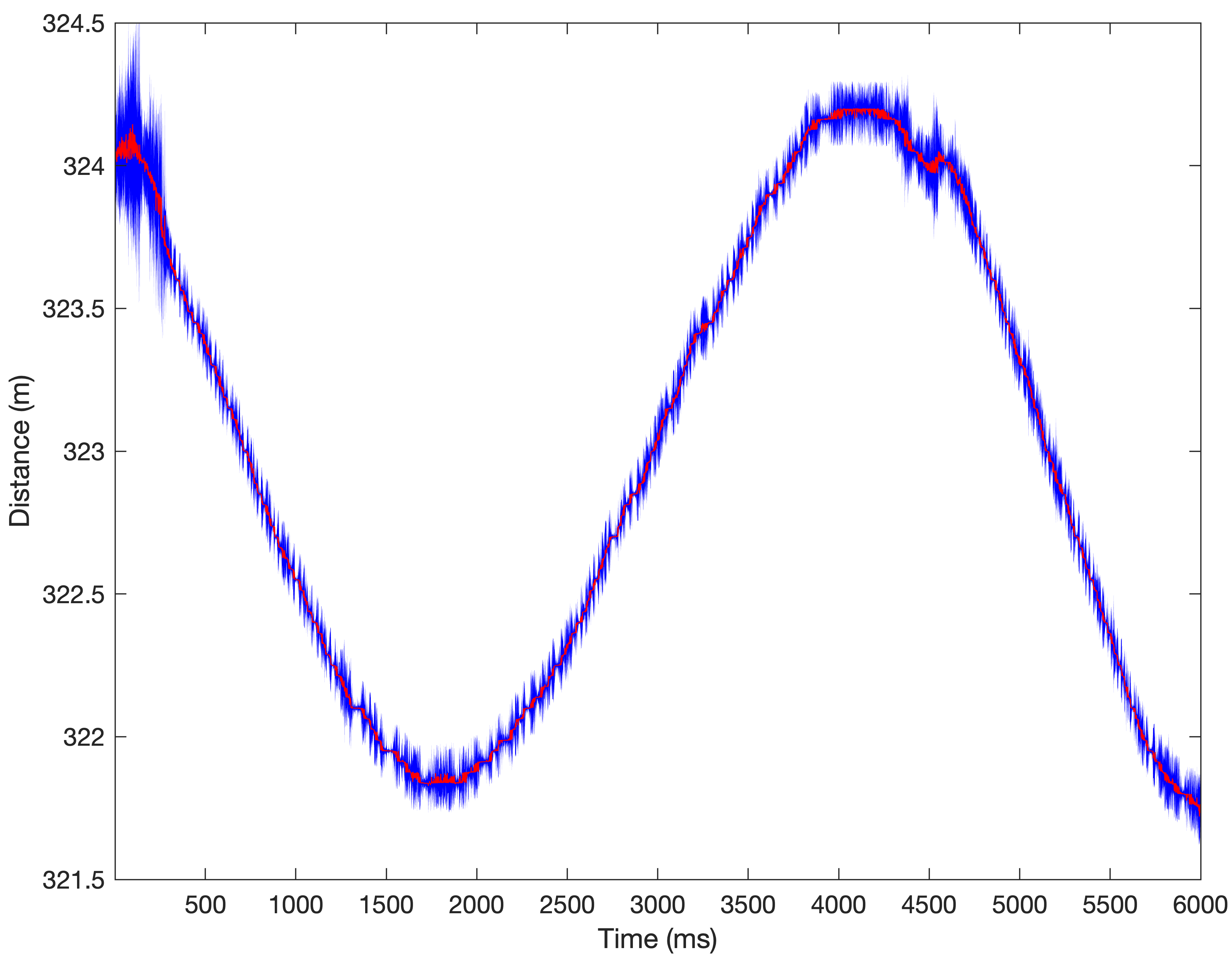}\\
	\caption{Temporal evolution of the estimated depth in pixel (16,16) of the second experiment (running pedestrian). The red curve represents the mean of the depth posterior distribution obtained for each frame and the blue region represents the $\pm 6$ standard deviation credible interval.}
	\label{fig:track_pedestrian}
\end{figure}

The second experiment was conducted under the same observation conditions and we recorded the movements of a pedestrian running back and forth at about 320 m from the detector. For this scene, we integrated the binary acquisitions into 1000 lidar frames per second. In each pixel and frame, we observe approximately 18 photons related to dark counts or ambient illumination from solar background, and the visible surfaces lead to 27 additional photons per pixel, on average. The proposed algorithm has been applied to a series of $6000$ successive frames, representing a total acquisition of $6$ s. As before, we set $\beta=0.5$, $M=5$ neighbors, $\sigma_{\textrm{RW}}=\sqrt{3}$ bins and $\nu_0=0.5$. The reconstructed point clouds, together with a standard video of the scene are presented in \href{https://youtu.be/A4cbl8DXOQk}{Video 5}.
Fig. \ref{fig:track_pedestrian} depicts the temporal profile of the depth estimated at a central pixel where the pedestrian is always visible. This figure shows the mean and credible interval ($\pm 6$ standard deviation interval for visualization purposes) of the depth posterior distribution. As expected, the depth uncertainty is larger at the beginning of the sequence due to the limit amount of information available about the object range. After about $300$ frames (300 ms), the uncertainty becomes more stable and the algorithm is able to successfully track the position of the surface. Note that such depth uncertainty measures can be used for instance to quantify uncertainties associated with the instantaneous velocity or moving objects.

Although the resolution of the $32\times32$ Princeton Lightwave Kestrel camera used in this study limits the resolution of the reconstructed point cloud, the proposed method can be applied to larger detector arrays without significant computational degradation since most computational steps can be performed at the pixel level, using only local information. Nonetheless, coarse depth maps, as obtained here, can be used in diverse applications involving object detection or recognition \cite{song2014sliding,bo2011depth}. Higher-resolution depth images can also be obtained by resorting to deep learning approaches such as in \cite{ni2017color,wen2018deep,Lindell:2018:3D}, where associated high resolution color maps are used as guidance to improve the details of the resulting fine depth map.

A last experiment was performed using synthetic data to demonstrate that the proposed method can be applied with larger SPAD arrays.
\begin{figure}[ht!]
	\center
	\includegraphics[width=\columnwidth]{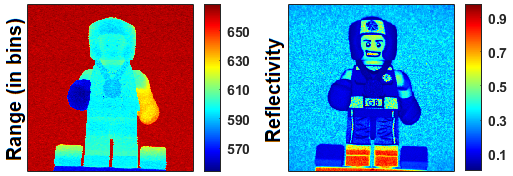}\\
	\caption{Depth (left) and reflectivity (right) profiles used to simulate the dynamic scene.}
	\label{fig:syntheticLego}
\end{figure}
The ground truth profiles used to simulate the synthetic data consists of $200\times 200$ pixels depth and reflectivity profiles obtained from real single-photon lidar data acquired at $532$nm in \cite{Tobin_SPIRAL}, with a target (a figurine) located at a distance of about 1.8 m from the imaging system. 
A dynamic scene of $312$ temporal frames was created from these profiles, such that the figurine enters in the field of view from the left with a horizontal shift of 1 pixel per frame (for the first $171$ frames). Once the figurine reaches the center of the image, a $45$ pixel-radius disk appears on the right hand-side of the image and is shifted to the left until it reaches the center of the image. Each pixel of the disk has a depth corresponding to $570$ bins and a reflectivity of $0.8$. 
Histograms of $1500$ temporal bins (bin width of $2$ps) were generated from \eqref{eq:lik_bg_poisson} using a Gaussian IRF with a FWHM of $31$ bins (same as that measured in \cite{Tobin_SPIRAL}), including a spatially constant background level, such that $MSC=35$ signal photons were recorded on average per pixel within each frame, with $SBR \approx 1.5$.
The proposed method has been applied using $M=5$, $\sigma_{\textrm{RW}}=\sqrt{3}$ bins, $\nu_0=0.5$ and $\beta=0.4$, even though using $\beta \in [0.4,0.6]$ did not change significantly the results. For visual comparison, a log-matched filtering depth estimation was also performed. Estimation of the background level was then conduced by averaging over the temporal bins outside the estimated peak (using the 3-sigma rule of thumb). Those estimates were finally plugged in \eqref{eq:lik_bg_poisson} to estimate the reflectivity profile by MLE.
The ground truth depth and reflectivity profiles, together with the estimated depth and reflectivity using both log-matched filtering and the proposed online approach are presented in \href{https://youtu.be/Ymdcn1CbNgg}{Video 6}. This video illustrates how the proposed spatio-temporal model regularizes the depth estimation process, while allowing new objects to enter the field of view.

\section{Conclusion}
\label{sec:conclusion}
In this work, we presented, to the best of our knowledge, a first algorithm for sequential reconstruction of dynamic 3D scenes from SPLs data, using temporal correlation. This method primarily focuses on the estimation of the surface depth using a model that does not involve the target reflectivity nor the background level. The resulting depth estimation process is particularly efficient, as is reduces to computing, pixel-wise, the cross-correlation (either discrete or continuous) between the measured photon ToAs and a modified IRF, which represents a fixed and predictable computational cost. Moreover, if the system IRF is Gaussian, the update rules are greatly simplified as the mean and variances of the pseudo-posterior distributions can be obtained in closed-form.
Thanks to the proposed spatio-temporal model, most of the steps of the algorithm can be performed independently for each pixel, which is attractive for parallel/distributed implementation. In this work, we focused of the theoretical development of the method but did not fully optimize its implementation, e.g., on GPUs, mainly because the current implementation of the detection method \cite{TachellaEUSIPCO2019} is not yet optimized. This is left for future work. We also derived the $\beta$-divergence for MSL data and the method proposed here could also be extended for online reconstruction of colored point clouds, provided that an efficient detection strategy (adapted to MSL data) is used. 

\appendix
\label{sec:appendix1}
In this appendix, we derive the expression of the $\beta$-divergence for the estimation of the depth $d$ in a single pixel and frame, assuming that multispectral single-photon lidar data (with $L$ wavelengths) are recorded simultaneously. We denote by $y_{\ell}$ the time of arrival of a photon at the $\ell$th wavelength. Using the classical MSL systems, separate and independent detectors are used such that the detection events in the $L$ channels are mutually independent, conditioned on the configuration of the scene. Thus, for any set of $L$ random variables $(y_{1},\ldots,y_{L})$ associated with each of the $L$ bands, the ideal model in \eqref{eq:lik_ideal} can be extended as 
\begin{eqnarray}
f_0(y_{1},\ldots,y_{L}|d)=\prod_{\ell=1}^L f_{\ell}(y_{\ell}|d)=\prod_{\ell=1}^L s_{\ell}\left(y_{\ell} - \frac{2 d}{c}\right),
\end{eqnarray}
where $s_{\ell}(\cdot)$ is the impulse response of the $\ell$th band \cite{altmann2017robust}. For multivariate continuous distributions the $\beta$-divergence has the same expression as in \eqref{eq:beta_div} and it can be easily shown in a similar fashion to \eqref{eq:beta_div2} that, 
under the same mild conditions as in Section \ref{sec:robust},\\
$D_{\beta}(\hat{f}(y_1,\ldots,y_L)||f_0(y_{1},\ldots,y_{L}|d))$
\begin{eqnarray}
\label{eq:div_MSL}
 & = & \textrm{Const.} -\dfrac{\beta+1}{\beta}\prod_{\ell=1}^L\dfrac{1}{K_{\ell}}\sum_{k}f_{\ell}(y_{\ell,k}|d)^{\beta}, 
\end{eqnarray}
where $\hat{f}(y_1,\ldots,y_L)$ is the product of the empirical marginal distributions of the ToAs in the $L$ bands, i.e., $\hat{f}(y_1,\ldots,y_L)=\prod_{\ell=1}^L\hat{f}(y_{\ell})$ and $\hat{f}(y_{\ell})=\dfrac{1}{K_{\ell}}\sum_{k=1}^{K_{\ell}}\delta(y_{\ell}-y_{\ell,k})$, 
with $\{y_{\ell,k}\}_k$ the ToAs of the $K_{\ell}$ photons detected at the $\ell$th
wavelength. Note that in contrast to classical maximum likelihood estimation which would introduce a sum (of the log-likelihood terms) over the $L$ bands, in \eqref{eq:div_MSL} we obtain a product over the $L$ bands and sums over the detection events. 

\begin{figure}[ht!]
	\center
	\includegraphics[width=\columnwidth]{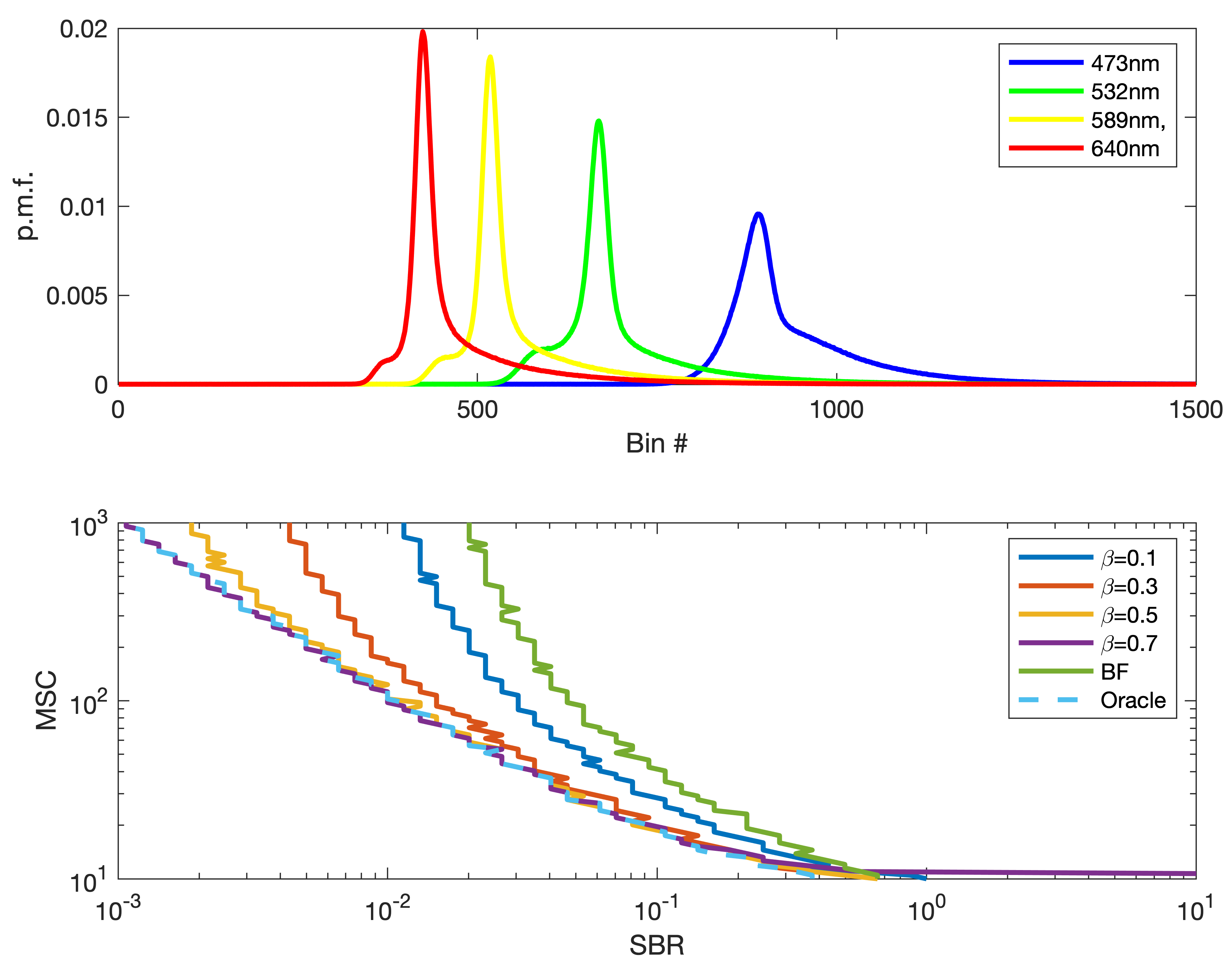}\\
	\vspace{-0.3cm}
	\caption{Top: IRFs from \cite{Tobin2017} used to simulate MSL synthetic data. Bottom: curves of satisfactory detection ($p_d=85\%$, with $\eta=28$) as a function of (SBR,MSC) using the real IRF. This figure compares only (pseudo-)MLEs adapted to MSL data.}
	\label{fig:MSL}
\end{figure}
To demonstrate the benefits of the proposed robust method for MSL-based depth imaging, we generated MSL data using the $L=4$ IRFs depicted in Fig. \ref{fig:MSL} (top), using the same SBR and same MSC for all the bands. In contrast to the results presented in Fig. \ref{fig:MMSE_synth}, for simplicity we only consider estimators using only the observations, without regularization or additional prior information, i.e., pseudo-MLEs. More precisely, we compared the minimum divergence estimator derived from \eqref{eq:div_MSL} to the BF MLE (assuming no background) for MSL data and the Oracle estimator assuming the reflectivity and background in each band is known. The limit of the regions of successful depth estimation ($p_d>85\%$ with $\eta=28$) are depicted in Fig. \ref{fig:MSL} (bottom). As in the single-wavelength case, increasing $\beta$ allows satisfactory depth estimation at lower SBRs than using BF and using $\beta=0.7$ here leads to results similar to those of the Oracle, without requiring knowledge or estimation of the reflectivity and background parameters. Note also that, as in the single-band case, too large values of $\beta$ (see $\beta=0.7$) lead to poor results in the low MSC and high SBR regime.

\section*{Acknowledgment}
This work was supported by the Royal Academy of Engineering under the Research Fellowship scheme RF201617/16/31, the ERC advanced grant C-SENSE, project 694888, the UK Defence Science and Technology Laboratory (DSTL X1000114765) and by the Engineering and Physical Sciences Research Council (EPSRC)  (grants EP/N003446/1, EP/T00097X/1 and EP/S000631/1) and the MOD University Defence Research Collaboration (UDRC) in Signal Processing. M. D. also acknowledges support from the Royal Society Wolfson Research Merit Award. We gratefully acknowledge the support of NVIDIA Corporation with the donation of the Titan Xp GPU used for this research.

\bibliography{mybibfile}
\bibliographystyle{IEEEtran}

\end{document}